\documentclass{article}%
\usepackage{amssymb}
\usepackage{amsfonts}
\usepackage{amsmath}%
\usepackage{graphicx}
\providecommand{\U}[1]{\protect\rule{.1in}{.1in}}
\newtheorem{theorem}{Theorem}

\newtheorem{corollary}[theorem]{Corollary}

\newtheorem{definition}[theorem]{Definition}
\newtheorem{example}[theorem]{Example}

\newtheorem{lemma}[theorem]{Lemma}

\newtheorem{proposition}[theorem]{Proposition}

\newenvironment{proof}[1][Proof]{\textbf{#1.} }{\ \rule{0.5em}{0.5em}}
\begin{document}

\title{Existence, uniqueness and efficiency of equilibrium in hedonic markets with
multidimensional types.}
\author{Ivar Ekeland\\Canada Research Chair in Mathematical Economics, UBC}
\date{First version, April 2005; this version, August 2008}
\maketitle

\begin{abstract}
We study equilibrium in hedonic markets, when consumers and suppliers have
reservation utilities, and the utility functions are separable with respect to
price. There is one indivisible good, which comes in different qualities; each
consumer buys $0$ or $1$ unit, and each supplier sells $0$ or $1$ unit.
Consumer types, supplier types and qualities can be either discrete of
continuous, in which case they are allowed to be multidimensional. Prices play
a double role: they keep some agents out of the market, and they match the
remaining ones pairwise. We define equilibrium prices and equilibrium
distributions, and we prove that equilibria exist, we investigate to what
extend equilibrium prices and distributions are unique, and we prove that
equilibria are efficient. In the particular case when there is a continuum of
types, and a generalized Spence-Mirrlees condition is satisfied, we prove the
existence of a pure equilibrium, where demand distributions are in fact demand
functions, and we show to what extent it is unique. The proofs rely on convex
analysis, and care has been given to illustrate the theory with examples.

\end{abstract}

\begin{center}

\end{center}

\section{Introduction.}

\subsection{Main results.}

In this paper, we show the existence and uniqueness of equilibrium in a
hedonic market, and we give uniqueness results. The main features of our model
are as follows:

\begin{itemize}
\item There is a single, indivisible, good in the market, and it comes in
different qualities $z$

\item Consumers and producers are price-takers and utility-maximizers. They
are characterized by the values of some variables; each set of values is
called a (multidimensional) \emph{type}.

\item Consumers buy at most one unit of the good, and they buy none if their
reservation utility is not met; producers supply at most one unit of the good,
and they supply none if their reservation utility is not met. In other words,
agents always have the option of staying out of the market.

\item The utilities of consumers and of producers are quasi-linear with
respect to price: the utility consumers with type $x$ derive from buying one
unit of quality $z$ at price $p\left(  z\right)  $ is $u\left(  x,z\right)
-p\left(  z\right)  $, and the utility producers with type $y$ derive from
selling one unit of quality $z$ at price $p\left(  z\right)  $ is $p\left(
z\right)  -v\left(  y,z\right)  $
\end{itemize}

Our results are valid in the discrete case and in the continuous case. We show
that there is a (nonlinear) price system $p\left(  z\right)  $ such that, for
every quality $z$, the number (or the aggregate mass) of consumers who demand
$z$ is equal to the number (or the aggregate mass) of suppliers who produce
$z.$ In addition, agents who are staying out of the market are doing so
because by entering they would lower their utility. In other words this price
system exactly\ matches a subset of consumers with a subset of producers, and
the remaining consumers or producers are priced out of the market. This is
called an \emph{equilibrium price}, and the resulting allocation of qualities
is called an \emph{equilibrium allocation}. An example is given in section
\ref{sec6}, and the reader may proceed there directly. We should stress,
however, that we prove existence in full generality, beyond the
one-dimensional situation described in that example.

Every price system $p\left(  z\right)  $ creates a matching between consumers
and producers: for every unit traded, there is a pair consisting of a consumer
who buys it and a producer who sells it. When summing their utilities, the
price of the traded item cancels out, so that the resulting utility of the
pair is independent of the price system. Unmatched consumers and producers
(singles) get their reservation utility. It is then meaningful to take the
social planner's point of view, and to ask for a matching between consumers
and producers which will maximize aggregate utility, where the utility of
matched pairs is the maximum utility they can get by trading, and the utility
of unmatched agents is their reservation utility. We will show that the
solution of this problem coincides with the equilibrium matching. This implies
that \emph{every equilibrium is efficient.}

An interesting feature of equilibrium pricing is that, even tough all
technologically feasible qualities are priced, not all of them will be traded
in equilibrium. For each non-traded quality, there is a non-empty bid-ask
range: all prices which fall within that range are equilibrium prices, that
is, they will not lure customers or suppliers away from traded qualities. This
means that equilibrium prices cannot be uniquely defined on non-traded
qualities. On the other hand, they are \emph{uniquely defined on traded
qualities}. There is a corresponding degree of uniqueness for the equilibrium allocation.

The main drawback of our model is the assumption that utilities are
quasi-linear. It is quite a restriction, from the economic point of view,
since it means that the marginal utility of money is constant, but our proof
seems to require it in an essential way. On the other hand, it also enables us
to prove some uniqueness results, which are probably not to be expected in the
more general case.

\subsection{The litterature.}

This paper inherits from two traditions in economics. On the one hand, it can
be seen as a contribution to the research program on hedonic pricing that was
outlined by Shervin Rosen in his seminal paper \cite{Rose}. The idea of
defining a good as a bundle of attributes (originating perhaps with Houthakker
\cite{Houtt}, and developed by Lancaster \cite{Lanc}, Becker \cite{Becker} and
Muth \cite{Muth}), provides a systematic framework for the economic analysis
of the supply and demand for quality. The main direction of investigation,
however, has been towards econometric issues, such as the construction of
price indices net of changes in quality; see for instance the seminal work of
Court \cite{Court} and the book \cite{Grili}). The identification of hedonic
models raises specific questions which have been first discussed by Rosen
\cite{Rose}, and most recently by Ekeland, Heckman and Nesheim \cite{Ekel}.
Theoretical question, such as the existence and characterization of
equilibria, have attracted less attention. The papers by Rosen \cite{Rose} and
later Mussa and Rosen \cite{Ros2} study the one-dimensional situation, that
is, the case when agents are fully characterized by the value of a single
parameter. The multidimensional situation has been investigated by Rochet and
Chon\'{e} \cite{Roc1}, but it deals with monopoly pricing. The issue of
equilibrium pricing in the multidimensional situation, had to my knowledge not
been adressed up to now (nor, for that matter, has the issue of oligopoly pricing).

One of Rosen's main achievement has been to recognize hedonic pricing as
nonlinear, against the prevailing tradition in econometric usage. As noted in
\cite{Rose}, a buyer can force prices to be linear with respect to quality if
certain types of arbitrage are allowed. In the present paper, buyers and
sellers are restricted to trading one unit of a single quality, and there is
no second-hand market, so this kind of arbitrage is unavailable, and prices
will be inherently nonlinear. This would not be the case if consumers and
producers were allowed to buy and sell several qualities simultaneously.

On the other hand, this paper also belongs to the tradition of assignment
problems. This tradition has several strands, one of which originates with
Koopmans and Beckmann \cite{Koop}, and the other with Shapley and Shubik
\cite{Shap}. We refer to the papers by Gretzki, Ostroy and Zame \cite{goz} and
\cite{gos2}, and to \cite{Rama} for more recent work. In this literature,
producers are not free to choose the quality they sell: each quality is
associated with a single producer, who can produce that one and not any other
one. The Shapley-Shubik model, for instance, describes a market for houses.
There are a certain number of sellers, each one is endowed with a house, and a
certain number of buyers. No seller can sell a house other than his own, but a
buyer can buy any house. This is in contrast with the situation in the present
paper, where both buyers and sellers are free to choose the quality they buy
or sell.

\subsection{Structure of the paper}

Section \ref{sec2} describes the mathematical model and the basic assumptions.
As we mentioned earlier, we do not require that the distribution of types be
continuous, nor that the number of consumers equals the number of producers.
Mathematically speaking, there is a positive measure $\mu$ on the set of
consumer types $X$, and a measure $\nu$ on the set of producer types $Y$, both
$\mu$ and $\nu$ can have atoms, and typically $\mu\left(  X\right)  \neq
\nu\left(  Y\right)  $. These features, although very appealing from the point
of view of economic modelling, introduce great complications in the
mathematical treatment. In earlier work \cite{IE}, the author has given a
streamlined proof in the particular case when $\mu$ and $\nu$ are non-atomic,
$\mu\left(  X\right)  =\nu\left(  Y\right)  $ and an additional sorting
assumption on utilities is satisfied (extending to multidimensional types the
classical Spence-Mirrlees single-crossing assumption), so that all agents with
the same type do the same thing. Beside the fact that it does not apply when
$X$ or $Y$ are finite, such a model does not capture one of the essential role
of prices, which serve not only to match consumers and producers which enter
the market (there must necessarily be an equal number of both) but also to
keep out of the market enough agents so that matching becomes possible. The
latter function is an essential focus of the present paper.

In our model, there is a single indivisible good, consumers are restricted to
buying one or zero unit, and producers are restricted to supply one or zero
unit. The price is a nonlinear function $p\left(  z\right)  $ of the quality
$z$. It is an equilibrium price if the market for every quality clears. This
implies that the number of consumers who trade is equal to the number of
suppliers who trade. The remaining, non-trading, agents, are kept out of the
market by the price system, which is either too high (for consumers) or too
low (for producer) to allow them to make more than their reservation utility.

It is important to note that in equilibrium consumers (or producers) which
have the same type may not be doing the same thing. This will typically occur
when utility maximisation does not result in a single quality being selected.
To be precise, given an equilibrium price $p\left(  z\right)  $, consumers of
type $x$ maximize $u\left(  x,z\right)  -p\left(  z\right)  $ with respect to
$z$. But there is no reason why there should be a unique optimal quality: even
if we assumed $u\left(  x,z\right)  $ to be strictly concave with respect to
$z$, the price $p\left(  z\right)  $ typically is nonlinear with respect to
$z$, and no conclusion can be derived about uniqueness.

If $p\left(  z\right)  $ is an equilibrium price, and if there is a
non-trivial subset $D\left(  x\right)  \subset Z$ such that any $z\in D\left(
x\right)  $ is a utility mazimizer for $x$, there will be a certain
equilibrium probability $P_{x}^{\alpha}$ on $D\left(  x\right)  $. This means
that, given $A\subset D\left(  x\right)  ,$ the number $P_{x}^{\alpha}\left[
A\right]  \in\left[  0,\ 1\right]  $ is the proportion of agents of type $x$
whose demands lie in $A$. Similarly, there will be an equilibrium probability
$P_{y}^{\beta}$ for every producer $y$, and the resulting demand and supply
for every quality $z$ will balance out. A formal definition is given in
section \ref{sec3}. In other words, in equilibrium, we cannot tell which agent
of a given type does what, but we can tell how many of them do this or that.

The main results of the paper, together with the definition of equilibrium,
are stated in section \ref{sec3}: equilibria exist, equilibrium prices are not
unique, there is a unique equilibrium allocation, and it is efficient (Pareto
optimal). Proofs are deferred to Appendices C and D. These proofs combine two
mathematical ingredients, the Hahn-Banach separation theorem on the one hand,
and duality techniques which extend the classical Fenchel duality for convex
functions, and which have been developed in the context of optimal
transportation (see \cite{Villa} for a recent survey). Everything relies in
studying a certain optimization problem (\ref{uni}), which is novel.

Section \ref{sec5} gives additional assumptions which ensure that all agents
of the same type do the same thing in equilibrium:\ $\mu$ and $\nu$ should be
non-atomic, and conditions (\ref{eq11}) and (\ref{eq12}) should be satisfied.
These conditions extend to multidimensional types the classical
single-crossing assumption of Spence and Mirrlees. The resulting equilibria
are called \emph{pure}, in reference to pure and mixed equilibria in game
theory. Note however that, even in this case, one cannot fully determine the
behaviour of agents in equilibrium: if consumers of type $x$ are indifferent
between entering the market or not (either decision giving them their
reservation utility), then, even with these additional assumptions, we cannot
say which ones will stay out and which ones will come in. The equilibrium
relations will only determine the proportion of each.

Subsection \ref{sec6} describes an explicit example. It is strictly
one-dimensional (types and qualities are real numbers), which makes
calculations possible, and a complete description of the equilibrium is
provided. Unfortunately, the method uses does not extend to multidimensional types.

Appendix A gives the mathematical results on $u$-convex and $v$-concave
analysis which will be in constant use in the text. Appendix B gives general
mathematical notations, and references about Radon measures. Appendices C and
D contain proofs.

\section{The model.\label{sec2}}

\subsection{Standing assumptions.}

Let $X\subset R^{d_{1}},Y\subset R^{d_{2}},$ and $Z_{0}\subset R^{d_{3}}$ be
compact subsets. We are given non-negative finite measures $\mu$ on $X$ and
$\nu$ on $Y.$ They are allowed to have point masses.

Typically, we will have $\mu\left(  X\right)  \neq\nu\left(  Y\right)  $.

Let $\Omega_{1}$ be a neighbourhood of $X\times Z_{0}$ in $R^{d_{1}+d_{3}}$,
and $\Omega_{2}$ be a neighbourhood of $Y\times Z_{0}$ in $R^{^{d_{2}+d_{3}}}%
$. We are given continuous functions $u:\Omega_{1}\rightarrow R$ and
$v:\Omega_{2}\rightarrow R$. It is assumed that $u$ is differentiable with
respect to $x$, and that the derivative:%
\[
D_{x}u=\left(  \frac{\partial u}{\partial x_{1}},...,\frac{\partial
u}{\partial x_{d_{1}}}\right)
\]
is continuous with respect to $\left(  x,z\right)  $. Similarly it is assumed
that $v$ is differentiable with respect to $y$, and that the derivative
$D_{y}v$ is continuous with respect to $\left(  y,z\right)  $.

Note that $X,Y$ and/or $Z_{0}$ are allowed to be finite. If $X$ is finite, the
assumption on $u$ is satisfied. If $Y\,\ $is finite, the assumption on $v$ is satisfied.

\subsection{Bid and ask prices}

We are describing the market for a quality good:\ it is indivisible, and units
differ by their characteristics $\left(  z_{1},...,z_{d_{3}}\right)  \in
Z_{0}$. The bundle $z=\left(  z_{1},...,z_{d_{3}}\right)  $ will be referred
to as a (multidimensional) \emph{quality}. So $Z_{0}$ is the set of all
technologically feasible qualities; it is to be expected that they will not
all be traded in equilibrium.

Points in $X$ represent consumer types, points in $Y$ represent producer
types. If $X\ $is finite, then $\mu\left(  x\right)  $ is the number of
consumers of type $x$. If $Y$ is finite, then $\nu\left(  y\right)  $ is the
number of producers of type $y$. If $X$ is infinite, then $\mu$ is the
distribution of types in the consumer population, and the same interpretation
holds for $\left(  Y,\nu\right)  $.

Each consumer buys zero or one unit, and each supplier sells zero or one unit.
There is no second-hand trade.

For the time being, we define a price system\emph{ }to be a continuous map
$p:Z_{0}\rightarrow R$. This definition will be modified in a moment, as the
set $Z_{0}$ will be extended to a larger set $Z$. Typically, pricing is
nonlinear with respect to the characteristics. Once the price system is
announced, agents make their decisions according to the following rules:

\begin{itemize}
\item Consumers of type $x$ maximize $u\left(  x,z\right)  -p\left(  z\right)
$ over $Z_{0}$. If the value of that maximum is strictly positive, the
consumer enters the market and buys one unit of the maximizing quality $z$. If
there are several maximizing qualities, he is indifferent between them, and
the way he chooses which one to buy is not specified at this stage. If the
value of the maximum is $0$, he is indifferent between staying out of the
market, and entering it to buy one unit of the maximizing quality. Again, the
way he chooses is not specified at this stage.

\item Producers of type $y$ maximize $p\left(  z\right)  -v\left(  y,z\right)
$ over $Z_{0}$. If the value of that maximum is strictly positive, the
producer enters the market and sells one unit of the maximizing quality $z$.
If there are several maximizing qualities, he is indifferent between them. If
the value of the maximum is $0$, he is indifferent between staying out of the
market, and entering it to sell one unit of the maximizing quality.
\end{itemize}

To model this procedure by a straigthforward maximization, we introduce two
extra points $\varnothing_{d}\notin Z_{0}$ and $\varnothing_{s}\notin Z_{0}$,
with $\varnothing_{d}$ $\neq\varnothing_{s}$, and we extend utilities and
prices as follows:%
\begin{align}
p\left(  \varnothing_{d}\right)   &  =u\left(  x,\varnothing_{d}\right)
=0\ \ \forall x\in X\label{w1}\\
p\left(  \varnothing_{s}\right)   &  =v\left(  y,\varnothing_{s}\right)
=0\ \ \forall y\in Y\label{w2}\\
u\left(  x,\varnothing_{s}\right)   &  =-1\ ,\ v\left(  y,\varnothing
_{d}\right)  =1 \label{w3}%
\end{align}

The set of possible decisions for agents is now
\[
Z=Z_{0}\cup\left\{  \varnothing_{d}\right\}  \cup\left\{  \varnothing
_{s}\right\}
\]
so that:%
\begin{align*}
\max\left\{  u\left(  x,z\right)  -p\left(  z\right)  \ |\ z\in Z\right\}   &
\geq u\left(  x,\varnothing_{d}\right)  -p\left(  \varnothing_{d}\right)  =0\\
\max\left\{  p\left(  z\right)  -v\left(  y,z\right)  \ |\ z\in Z\right\}   &
\geq p\left(  \varnothing_{s}\right)  -v\left(  y,\varnothing_{s}\right)  =0
\end{align*}
and the procedure we just described amounts to maximizing over $Z$ instead of
$Z_{0}$. The relations (\ref{w1}) to (\ref{w3}) imply that consumers will
never choose $\varnothing_{s}$ (it is always better to choose $\varnothing
_{d}$), and producers will never choose $\varnothing_{d}$ (it is always better
to choose $\varnothing_{s}$). So our model does capture the intended behaviour.

Note that we have normalized reservation utilities to $0$. This does not cause
any loss of generality. The behaviour of consumers, for instance, is fully
specified by $u\left(  x,z\right)  $ and $\bar{u}\left(  x\right)  $, the
latter being the reservation utility, and we get the same behaviour by
replacing $u\left(  x,z\right)  $ by $u\left(  x,z\right)  -\bar{u}\left(
x\right)  $ and $\bar{u}\left(  x\right)  $ by $0$, the only restriction being
that we would require $\bar{u}$ to be $C^{1}$, to preserve the regularity
properties of $u$.

Normalizing reservation utilities to $0$, we find that $u\left(  x,z\right)  $
is the bid price for quality $z$ by consumers of type $x$, that is, the
highest price that they are willing to pay for that quality. Similarly,
$v\left(  y,z\right)  $ is the asking price for quality $z$ by producers of
type $y\,$, that is, the lowest price they are willing to accept for supplying
that quality. For a given quality $z\in Z$, it is natural to consider the
highest bid price from consumers and the lowest ask price from producers:

\begin{definition}
The \emph{highest bid price} $b:Z\rightarrow R$ is given by:%
\[
b\left(  z\right)  =\max_{x}u\left(  x,z\right)
\]
and the \emph{lowest ask price} $a:Z\rightarrow R$ is given by:%
\[
a\left(  z\right)  =\min_{y}v\left(  y,z\right)
\]

\end{definition}

Note that $\ b\left(  \varnothing_{d}\right)  =a\left(  \varnothing
_{s}\right)  =0$ and that $a\left(  \varnothing_{d}\right)  =-b\left(
\varnothing_{s}\right)  =1$.

It follows from their definitions that $b$ is $u$-convex and $a$ is
$v$-concave. More precisely, we have $b\left(  z\right)  =0_{x}^{\sharp}$ and
$a\left(  z\right)  =0_{y}^{\flat}$ where $0_{x}$ and $0_{y}$ denote the maps
$x\rightarrow0$ and $y\rightarrow0$ on $X$ and $Y$. Conversely, we have
$0=\max_{z}\left\{  u\left(  x,z\right)  -b\left(  z\right)  \right\}  $ and
$0=\min_{z}\left\{  v\left(  y,z\right)  -a\left(  z\right)  \right\}  $, so
that $b^{\sharp}\left(  x\right)  =0$ and $a^{\flat}\left(  y\right)  =0$.

Note that if the price system is such that $p\left(  z\right)  >b\left(
z\right)  $ for some quality $z$, then there will be no buyers for this
quality, and so it cannot be traded at that price. Similarly, if $p\left(
z\right)  <a\left(  z\right)  $, then there will be no sellers for this
quality, and it cannot be traded at that price. The following is obvious:

\begin{proposition}
[No-trade equilibrium]If $a\left(  z\right)  >b\left(  z\right)  $ everywhere,
then all consumers and all producers stay out of the market.
\end{proposition}

\subsection{Demand and supply}

From now on, a \emph{price system} will be a continuous map $p:Z\rightarrow R$
such that $p\left(  \varnothing_{d}\right)  =p\left(  \varnothing_{s}\right)
=0$.

Given a price system $p$, the map $p:Z\rightarrow R$ is continuous and the set
$Z$ is compact, so that the functions $u\left(  x,z\right)  -p\left(
z\right)  $ and $p\left(  z\right)  -v\left(  y,z\right)  $ attain their
maximum on $Z$.

\begin{definition}
Given a price system $p$, we define:%
\begin{align*}
D\left(  x\right)   &  =\arg\max\left\{  u\left(  x,z\right)  -p\left(
z\right)  \ |\ z\in Z\right\} \\
S\left(  y\right)   &  =\arg\min\left\{  v\left(  y,z\right)  -p\left(
z\right)  \ |\ z\in Z\right\}
\end{align*}
Both are compact and non-empty subsets of $Z$. We shall refer to $D\left(
x\right)  $ as the \emph{demand} of type $x$ consumers, and to $S\left(
y\right)  $ as the \emph{supply} of type $y$ producers.
\end{definition}

It follows from the definitions that if a consumer of type $x$ is out of the
market, then we must have $\varnothing_{d}\in$ $D\left(  x\right)  .$ If there
is no other point in $D\left(  x\right)  $, then all consumers of the same
type stay out of the market. If, on the other hand, $D\left(  x\right)  $
contains some point $z\in Z_{0}$, then all consumers of type $x$ are
indifferent between staying out or buying quality $z$, and we may expect that
some of them actually buy quality $z$ instead of staying out. This remark will
be at the core of our equilibrium analysis. Of course, the same observation is
valid for producers.

The following result clarifies the relation between $D\left(  x\right)  $ and
$S\left(  y\right)  $ on the one hand, and the sub- and supergradients
$\partial p^{\sharp}\left(  x\right)  $ and $\partial p^{\flat}\left(
y\right)  $ on the other. Recall that:%
\begin{align*}
p^{\sharp}\left(  x\right)   &  =\max\left\{  u\left(  x,z\right)  -p\left(
z\right)  \ |\ z\in Z\right\} \\
p^{\flat}\left(  y\right)   &  =\min\left\{  v\left(  y,z\right)  -p\left(
z\right)  \ |\ z\in Z\right\}
\end{align*}

\begin{proposition}
\label{p6}We have $D\left(  x\right)  \subset\partial p^{\sharp}\left(
x\right)  $ and $S\left(  y\right)  \subset\partial p^{\flat}\left(  y\right)
$. More precisely:%
\begin{align*}
D\left(  x\right)   &  =\left\{  z\in\partial p^{\sharp}\left(  x\right)
\ |\ p\left(  z\right)  =p^{\sharp\sharp}\left(  z\right)  \right\} \\
S\left(  y\right)   &  =\left\{  z\in\partial p^{\flat}\left(  y\right)
\ |\ p\left(  z\right)  =p^{\flat\flat}\left(  z\right)  \right\}
\end{align*}

\end{proposition}

\begin{proof}
The point $x\in X$ being fixed, consider the functions $\varphi:Z\rightarrow
R$ and $\psi:Z\rightarrow R$ defined by $\varphi\left(  z\right)  =u\left(
x,z\right)  -p\left(  z\right)  $ and $\psi\left(  z\right)  =u\left(
x,z\right)  -p^{\sharp\sharp}\left(  z\right)  $. The subgradient $\partial
p^{\sharp}\left(  x\right)  $ is the set of points $z$ where $\psi$ attains
its maximum (see appendix \ref{sec1}), while $D\left(  x\right)  $ is the set
of points $z$ where $\varphi$ attains its maximum. But $\psi\geq\varphi$ and
$\max\psi=\max\varphi$. The result follows.
\end{proof}

\begin{definition}
Given a price system $p\left(  z\right)  $, consumers of type $x$ are
\emph{inactive} if $p^{\sharp}\left(  x\right)  <0$, so that $D\left(
x\right)  =\left\{  \varnothing_{d}\right\}  $, and they are \emph{active} if
$p^{\sharp}\left(  x\right)  >0$, so that $\left\{  \varnothing_{d}\right\}
\notin D\left(  x\right)  $. They are \emph{indifferent} if $p^{\sharp}\left(
x\right)  =0$, so that $D\left(  x\right)  \supset\left\{  \varnothing
_{d}\right\}  \cup\left\{  z\right\}  $ for some $z\in Z_{0}$. Similarly,
producers of type $y$ are inactive, active or indifferent according to whether
$p^{\flat}\left(  y\right)  $ is positive, negative or zero.
\end{definition}

\subsection{Admissible price systems}

We have seen that, if $a\left(  z\right)  >b\left(  z\right)  $ everywhere,
there is a no-trade equilibrium. We are concerned with the more interesting
case when $a\left(  z\right)  \leq b\left(  z\right)  $ for some $z.$

\begin{definition}
Quality $z\in Z$ is \emph{marketable} if $a\left(  z\right)  \leq b\left(
z\right)  $. The set of marketable qualities will be denoted by $Z_{1}:$
\begin{align*}
Z_{1}  &  =\left\{  z\in Z\;|\;a\left(  z\right)  \leq b\left(  z\right)
\right\} \\
&  =\left\{  z\in Z\;|\;\exists\,x\;,\exists\,y:v\left(  y,z\right)
\leq\;u\left(  x,z\right)  \right\}
\end{align*}

\end{definition}

Note that staying out is not a marketable option: $a\left(  \varnothing
_{d}\right)  >b\left(  \varnothing_{d}\right)  $ and $a\left(  \varnothing
_{s}\right)  >b\left(  \varnothing_{s}\right)  $. As mentioned earlier, this
means that consumers will never choose $\varnothing_{s}$ and that suppliers
will never choose $\varnothing_{d}$. We have therefore the inclusions:
\[
Z_{1}\subset Z_{0}\varsubsetneq Z
\]

If a quality $z$ is not marketable, one will never be able to find a
buyer/seller pair that trade $z$. If a quality $z$ is marketable, there is no
sense in setting its price to be higher than $b\left(  z\right)  $ (there
would be no buyers), or lower than $a\left(  z\right)  ~$(there would be no
sellers). Hence:

\begin{definition}
A\emph{\ }price system $p:Z\rightarrow R$ will be called \emph{admissible}
if:
\[
\forall z\in Z_{1},\;\;a\left(  z\right)  \leq p\left(  z\right)  \leq
b\left(  z\right)
\]

\end{definition}

Let $p$ be an admissible price system, so that $a\left(  z\right)  \leq
p\left(  z\right)  \leq b\left(  z\right)  $. Recall that $p^{\sharp}\left(
x\right)  $ is the indirect utility of type $x$ consumers, and that
$-p^{\flat}\left(  y\right)  $ is the indirect utility of type $y$ producers.
Taking conjugates, we get:
\begin{align*}
\forall x  &  \in X,\;\;0\leq p^{\sharp}\left(  x\right) \\
\forall y  &  \in Y,\ \ \ \ 0\geq p^{\flat}\left(  y\right)
\end{align*}
which means that all consumers and producers achieve at least their
reservation utility.

\section{Equilibrium\label{sec3}}

\subsection{Demand distribution and supply distribution}

Assume a price system $p:Z\rightarrow R$ is given. Let $D\left(  x\right)  $
and $S\left(  y\right)  $ be the associated demand and supply. Recall. that
their graphs are compact sets.

We refer to Appendix B for notations and definitions concerning Radon measures
and probabilities.

\begin{definition}
A \emph{demand distribution} associated with $p$ is a positive measure
$\alpha_{X\times Z}$ on $X\times Z$ such that:

\begin{itemize}
\item $\alpha_{X\times Z}$ is carried by the graph of $D$

\item its marginal $\alpha_{X}$ is equal to $\mu$
\end{itemize}

Similarly, a \emph{supply distribution} associated with $p$ is a positive
measure $\beta_{Y\times Z}$ on $Y\times Z$ such that:

\begin{itemize}
\item $\beta_{Y\times Z}$ is carried by the graph of $S$

\item its marginal $\beta_{Y}$ is equal to $\nu$
\end{itemize}
\end{definition}

The conditional probabilities $P_{x}^{\alpha}$ and $P_{y}^{\beta}$ then are
carried by $D\left(  x\right)  $ and $S\left(  y\right)  $ respectively. Given
$A\subset Z$, the numbers $P_{x}^{\alpha}\left[  A\right]  $ and $P_{y}%
^{\beta}\left[  A\right]  $ are readily interpreted as the probability that
consumers of type~$x$ demand some $z\in A$ and the probability that producers
of type $y$ supply some $z\in A$.

If $S\left(  y\right)  $ is a singleton, so that the supply of type $y$
producers is uniquely defined, then $P_{y}^{\beta}$ reduces to a Dirac mass:%
\[
S\left(  y\right)  =\left\{  s\left(  y\right)  \right\}  \Longrightarrow
P_{y}^{\beta}=\delta_{s\left(  y\right)  }%
\]
and similarly for consumers.

\subsection{Definition of equilibrium}

\begin{definition}
An \emph{equilibrium} is a triplet $(p,\alpha_{X\times Z},\beta_{Y\times Z})$,
where $p$ is an admissible price system and $\alpha_{X\times Z}$ and
$\beta_{Y\times Z}$ are demand and supply distributions associated with $p$,
such that:%
\[
\alpha_{Z_{0}}=\beta_{Z_{0}}%
\]

\end{definition}

By $\alpha_{Z_{0}}$ and $\beta_{Z_{0}}$ we denote the marginals of
$\alpha_{X\times Z}$ and $\beta_{Y\times Z}$ on $Z_{0}$. Let us write down
explicitly all the conditions on $(p,\alpha,\beta)$ implied by this definition:

\begin{enumerate}
\item $p:Z\rightarrow R$ is continuous, and $p\left(  z\right)  \in\left[
a\left(  z\right)  ,\ b\left(  z\right)  \right]  $ whenever $a\left(
z\right)  \leq b\left(  z\right)  $

\item the marginal $\alpha_{X}$ is equal to $\mu$

\item the conditional probability $P_{x}^{\alpha}$ is carried by $D\left(
x\right)  $

\item the marginal $\beta_{Y}$ is equal to $\nu$

\item the conditional probability $P_{y}^{\beta}$ is carried by $S\left(
y\right)  $

\item the marginals $\alpha_{Z}$ and $\beta_{Z}$ coincide on $Z_{0}$:%
\[
\alpha_{Z}\left[  A\right]  =\beta_{Z}\left[  A\right]  \ \ \ \forall A\subset
Z_{0}%
\]

\end{enumerate}

The interpretation is as follows. Given $p$, consumers of type $x$ maximize
their utility, thereby defining their individual demand set $D\left(
x\right)  $. If that set is a singleton, $D\left(  x\right)  =\left\{
d\left(  x\right)  \right\}  $, the probability $P_{x}^{\alpha}$ must be the
Dirac mass carried by $d\left(  x\right)  $, and all consumers of type $x$ do
the same thing: they stay out of the market if $d\left(  x\right)
=\varnothing_{d}$, and they buy $z\in Z_{0}$ if $d\left(  x\right)  =z$. If
$D\left(  x\right)  $ contains several points, then consumers of type $x$ are
indifferent among these alternatives, and they all do different things. For
any Borel subset $A\subset$ $D\left(  x\right)  $, the probability\ $P_{x}%
^{\alpha}\left[  A\right]  $ gives us the proportion of consumers of type $x$
who choose some $z\in A$ in equilibrium.

Similar considerations hold for suppliers. Condition 6 just states that
markets clear in equilibrium:\ for every quality $z\in Z_{0}$, the number (or
the aggregate mass) of buyers equals the number (or the aggregate mass) of
suppliers. Note that this number (or this mass) might be zero, meaning that
this particular quality is not traded. This will happen, for instance, if
$a\left(  z\right)  >b\left(  z\right)  $, so that quality $z$ is not
marketable. It follows that, in equilibrium, demand and supply are carried by
$Z_{1}$, the set of marketable qualities:%
\[
\alpha_{Z}\left[  Z_{1}\right]  =\alpha_{Z}\left[  Z_{0}\right]  =\beta
_{Z}\left[  Z_{0}\right]  =\beta_{Z}\left[  Z_{1}\right]
\]

The number (or the aggregate mass) of consumers who stay out of the market is
$\alpha_{Z}\left(  \left\{  \varnothing_{d}\right\}  \right)  $, and the
number (or the aggregate mass) of producers who stay out of the market is
$\beta_{Z}\left(  \left\{  \varnothing_{s}\right\}  \right)  $. As we
mentioned several times before, we must have $\alpha_{Z}\left(  \left\{
\varnothing_{s}\right\}  \right)  =0$ and $\beta_{Z}\left(  \left\{
\varnothing_{d}\right\}  \right)  =0$.

\subsection{Main results}

We begin by an existence result:

\begin{theorem}
[Existence]Under the standing assumptions, there is an equilibrium.
\end{theorem}

As noted above, if the set $Z_{1}$ of marketable qualities is empty, there is
an equilibrium, namely the no-trade equilibrium, and it is unique. From now on
we assume $Z_{1}\neq\varnothing$. The Existence Theorem will be proved in
section \ref{sec20}.

There is no uniqueness of equilibrium prices. For instance, if a quality $z\in
Z_{0}$ is non-marketable, its price $p\left(  z\right)  \ $can be specified
arbitrarily. More generally, in section \ref{sec20} we will prove the
following (see Proposition \ref{p11}):

\begin{theorem}
[Non-uniqueness of equilibrium prices]The set of all equilibrium prices $p$ is
convex and non-empty. If $p:Z\rightarrow R$ is an equilibrium price, then so
is every $q:Z\rightarrow R$ which is admissible, continuous, and satisfies:%
\begin{equation}
p^{\sharp\sharp}\left(  z\right)  \leq q\left(  z\right)  \leq p^{\flat\flat
}\left(  z\right)  \text{ \ \ }\forall z\in Z \label{mar}%
\end{equation}
For $\alpha$- and $\beta$-almost every quality $z$ which is traded in
equilibrium, we have $p^{\sharp\sharp}\left(  z\right)  =p\left(  z\right)
=p^{\flat\flat}\left(  z\right)  $.
\end{theorem}

Note that $q$ is also required to be admissible, so that in addition to
(\ref{mar}) it has to satisfy the inequality:%
\begin{equation}
a\leq q\leq b \label{mark}%
\end{equation}

The economic interpretation is as follows. If $\left(  p,\alpha_{X\times
Z},\beta_{Y\times Z}\right)  $ \ is an equilibrium, there will be qualities
$z$ which are marketable, but which are not traded in equilibrium, because
every supplier type $y$ and every consumer type $x$ prefers some other
quality, which means that the price $p\left(  z\right)  $ is too low to
interest suppliers, and too high to interest consumers. Formulas (\ref{mar})
and (\ref{mark}) give the range of prices for which this situation will
persist. As long as the price $p\left(  z\right)  $ stays in the open
interval
\[
]\max\left\{  a\left(  z\right)  ,\ p^{\sharp\sharp}\left(  z\right)
\right\}  ,\min\left\{  b\left(  z\right)  ,\ p^{\flat\flat}\left(  z\right)
\right\}  [
\]
the quality $z$ will not be traded. In other words, the price of non-traded
qualities can be changed, within a certain range, without affecting
$\alpha_{X\times Z}$ or $\beta_{Y\times Z}$, that is, the equilibrium
distribution of consumers and suppliers. This is the major source of
non-uniqueness in equilibrium prices. On the other hand, if a quality $z$ is
traded in equilibrium, one cannot change the price $p\left(  z\right)  $
without affecting $\alpha_{X\times Z}$ and $\beta_{Y\times Z},$ that is,
without destroying the given equilibrium.

The equilibrium price $p$ is not unique, but the following result shows that
the demand and supply maps $D\left(  x\right)  $ and $S\left(  y\right)  $
almost are:

\begin{theorem}
[Quasi-uniqueness of equilibrium allocations]\label{12}Let $\left(
p_{1},\alpha_{X\times Z}^{1},\beta_{Y\times Z}^{1}\right)  $ and $\left(
p_{2},\alpha_{X\times Z}^{2},\beta_{Y\times Z}^{2}\right)  $ be two
equilibria. Denote by $D_{1}\left(  x\right)  ,D_{2}\left(  x\right)  $ and
$S_{1}\left(  y\right)  ,S_{2}\left(  y\right)  $ the corresponding demand and
supply maps. Denote by $P_{x}^{1},P_{y}^{1}$ and $P_{x}^{2},P_{y}^{2}$ the
corresponding conditional probabilities of demand and supply. Then:%
\begin{align*}
P_{x}^{2}\left[  D_{1}\left(  x\right)  \right]   &  =P_{x}^{1}\left[
D_{1}\left(  x\right)  \right]  =1\text{ \ for }\mu\text{-a.e. }x\\
P_{y}^{2}\left[  S_{1}\left(  y\right)  \right]   &  =P_{y}^{1}\left[
S_{1}\left(  y\right)  \right]  =1\text{ \ for }\nu\text{-a.e. }y
\end{align*}

\end{theorem}

In other words, any $z$ which types $x$ demands in the second equilibrium,
when prices are $p_{2}$, must belong to the demand set of $x$ when prices are
$p_{1}$ (even though $x$ might not demand it in the second equilibrium)

\begin{corollary}
If the demand of consumers of type $x$ is single-valued in the first
equilibrium, $D_{1}\left(  x\right)  =\left\{  d_{1}\left(  x\right)
\right\}  $, then $d_{1}\left(  x\right)  \in D_{2}\left(  x\right)  $. If
their demand is single-valued in the second equilibrium as well, then
$d_{1}\left(  x\right)  =d_{2}\left(  x\right)  $.
\end{corollary}

\begin{proof}
We have $P_{x}^{2}\left[  d_{1}\left(  x\right)  \right]  =1=P_{x}^{2}\left[
D_{2}\left(  x\right)  \right]  $. So $d_{1}\left(  x\right)  $ must belong to
$D_{2}\left(  x\right)  $, and the remainder$\ $must have zero probability:
\[
P_{x}^{2}\left[  D_{2}\left(  x\right)  \diagdown\left\{  d_{1}\left(
x\right)  \right\}  \right]  =0
\]

\end{proof}

\begin{corollary}
Let $\left(  p_{1},\alpha_{X\times Z}^{1},\beta_{Y\times Z}^{1}\right)  $ and
$\left(  p_{2},\alpha_{X\times Z}^{2},\beta_{Y\times Z}^{2}\right)  $ be two
equilibria. If consumers of type $x$ are inactive in the first equilibrium,
they cannot be active in the second.
\end{corollary}

\begin{proof}
Since $D_{1}\left(  x\right)  =\left\{  \varnothing_{d}\right\}  $, we must
have $\varnothing_{d}\in D_{2}\left(  x\right)  $. Assume consumers of type
$x$ are active in the second equilibrium. We must have $u\left(  x,z\right)
-p\left(  z\right)  >0$ for all $z\in D_{2}\left(  x\right)  $, including
$z=\varnothing_{d}$. Since $u\left(  x,\varnothing_{d}\right)  =p\left(
\varnothing_{d}\right)  =0$, this is a contradiction.
\end{proof}

Finally, we will show that we can find equilibrium demand and supply as
solutions of the planner's problem. With every pair of demand and supply
distributions, $\alpha_{X\times Z}^{\prime}$ and $\beta_{Y\times Z}^{\prime}$,
we associate the number:%

\begin{align*}
J\left(  \alpha_{X\times Z}^{\prime},\beta_{Y\times Z}^{\prime}\right)   &
=\int_{X\times Z}u\left(  x,z\right)  d\alpha_{X\times Z}^{\prime}%
-\int_{Y\times Z}v\left(  y,z\right)  d\beta_{Y\times Z}^{\prime}\\
&  =\int_{X}P_{x}^{\alpha^{\prime}}\left[  u\left(  x,z\right)  \right]
d\mu\left(  x\right)  -\int_{Y}P_{y}^{\beta^{\prime}}\left[  v\left(
y,z\right)  \right]  d\nu\left(  y\right)
\end{align*}

Note that all expectations are taken over $Z=Z_{0}\cup\left\{  \varnothing
_{d}\right\}  \cup\left\{  \varnothing_{s}\right\}  $. For a given $x$, the
first one $E_{x}^{\alpha^{\prime}}\left[  u\left(  x,z\right)  \right]  $
represents the average utility of consumers of type $x$. If they are all out
of the market, this average utility is zero, if some of them are out and
others in, the contribution of those who are out is zero. Similarly, the
second one $E_{y}^{\beta^{\prime}}\left[  v\left(  y,z\right)  \right]  $
represents the average cost of producers of type $y$. The sum $J$ therefore is
the aggregate utility of society resulting from $\alpha_{X\times Z}^{\prime}$
and $\beta_{Y\times Z}^{\prime}$ consumers and suppliers being equally weighted.

In the following, we restrict attention to demand and supply distributions
$\alpha_{X\times Z}^{\prime}$ and $\beta_{Y\times Z}^{\prime}$ such that the
marginals $\alpha_{Z_{0}}^{\prime}$ and $\beta_{Z_{0}}^{\prime}$ are equal.
These are the only ones that are relevant to the planner's problem, which
consists of matching producers and consumers so as to maximize social surplus.
The solution to that problem turns out to be precisely the equilibrium allocation.

\begin{theorem}
[Pareto optimality of equilibrium allocations]Let $\left(  p,\alpha_{X\times
Z},\beta_{Y\times Z}\right)  $ be an equilibrium. Take any pair of demand and
supply distributions $\alpha_{X\times Z}^{\prime}$ and $\beta_{Y\times
Z}^{\prime}$ such that $\alpha_{Z_{0}}^{\prime}=\beta_{Z_{0}}^{\prime}$. Then
\begin{equation}
J\left(  \alpha_{X\times Z}^{\prime},\beta_{Y\times Z}^{\prime}\right)  \leq
J\left(  \alpha_{X\times Z},\beta_{Y\times Z}\right)  =\int_{X}p^{\sharp
}\left(  x\right)  d\mu-\int_{Y}p^{\flat}\left(  y\right)  d\nu\label{par}%
\end{equation}

\end{theorem}

The proof of the two last theorems will be given in section \ref{sec21}.

\subsection{Example 1: the case of a single quality.}

Let $Z_{0}=\left\{  z\right\}  $. In other words, there is a single
technologically feasible quality. While this example does not have great
economic interest, it is quite illuminating to see what the various
assumptions mean and how the preceding results apply to this case.

We introduce $Z=\left\{  z\right\}  \cup\left\{  \varnothing_{d}\right\}
\cup\left\{  \varnothing_{s}\right\}  $. \ For the sake of simplicity,
consider the case when $X$ and $Y$ are finite. Set $u\left(  x,z\right)
=u\left(  x\right)  $ and $v\left(  y,z\right)  =v\left(  y\right)  $ and
$p\left(  z\right)  =p$. Indirect utilities are given by:%
\begin{align*}
\max\left\{  u\left(  x\right)  -p,0\right\}   &  =p_{x}^{\sharp}\text{ for
}x\\
\max\left\{  p-v\left(  y\right)  ,0\right\}   &  =-p_{y}^{\flat}\text{ for }y
\end{align*}

The highest bid price for $z$ is $b=\max_{x}u\left(  x\right)  $, and the
lowest ask price is $a=\min_{y}v\left(  y\right)  $.

If $b<a$, then the quality $z$ is not marketable, and the no-trade equilibrium prevails.

Suppose $b\geq a$. A price $p$ is admissible if $a\leq p\leq b$. Set:%
\begin{align*}
I_{1}\left(  p\right)   &  =\left\{  x\in X\ |\ u\left(  x\right)
<p\ \right\} \\
I_{2}\left(  p\right)   &  =\left\{  x\in X\ |\ u\left(  x\right)
=p\ \right\} \\
I_{3}\left(  p\right)   &  =\left\{  x\in X\ |\ u\left(  x\right)
>p\ \right\}
\end{align*}
and define $J_{1}\left(  p\right)  ,J_{2}\left(  p\right)  ,J_{3}\left(
p\right)  $ in a similar way for producers. An equilibrium is a set $\left(
p,\alpha,\beta\right)  $ such that

\begin{itemize}
\item $\alpha=\left(  \alpha_{x}\right)  ,\ x\in X$, where each $\alpha_{x}$
is a probability on $\left\{  z\right\}  \cup\left\{  \varnothing_{d}\right\}
$

\item $\beta=\left(  \beta_{y}\right)  ,\ y\in Y,$ where each $\beta_{y}$ is a
probability on $\left\{  z\right\}  \cup\left\{  \varnothing_{s}\right\}  $

\item $\sum_{x}\alpha_{x}\left(  z\right)  =\sum_{y}\beta_{y}\left(  z\right)
$
\end{itemize}

Let us translate this. If $x\in I_{1}\left(  p\right)  $, then consumers of
type $x$ stay out of the market, so that $\alpha_{x}\left(  z\right)  =0$. If
$x\in I_{3}\left(  p\right)  $, then consumers of type $x$ buy $z$, so that
$\alpha_{x}\left(  z\right)  =1$. If $i\in I_{2}\left(  p\right)  ,$ then
$\alpha_{x}\left(  z\right)  $ is the proportion of consumers of type $x$ who
buy $z$ in equilibrium. Denote by $\#\left[  A\right]  $ the number of
elements in a finite set $A$. The equilibrium condition implies that:%
\begin{align}
\#\left[  I_{3}\left(  p\right)  \right]   &  \leq\#\left[  J_{2}\left(
p\right)  \cup J_{3}\left(  p\right)  \right] \label{in11}\\
\#\left[  J_{3}\left(  p\right)  \right]   &  \leq\#\left[  I_{2}\left(
p\right)  \cup I_{3}\left(  p\right)  \right]  \label{in12}%
\end{align}

Conversely, if these two inequalities are satisfied, we will always be able to
find numbers $\alpha_{x}$ and $\beta_{y}$ such that $0\leq\alpha_{x}\leq1$,
$\alpha_{x}=0$ if $\ x\in I_{1}\left(  p\right)  $ and $\alpha_{x}=1$ if $x\in
I_{3}\left(  p\right)  $, with corresponding constraints for the $\beta_{y}$.
So, in that particular case, the equilibrium conditions boil down to the
inequalities (\ref{in11}) and (\ref{in12}).

Note that there is no uniqueness of the equilibrium price $p$. \ If for
instance $u_{\bar{x}}>v_{\bar{y}}$, with $u_{x}<v_{\bar{y}}$ for all
$x\neq\bar{x}$ and $v_{y}>u_{\bar{x}}$ for all $y\neq\bar{y}$, then any price
$p\in\left[  u_{\bar{x}},\ v_{\bar{y}}\right]  $ is an equilibrium price.
There is no uniqueness of the equilibrium allocation either. If for instance
$u_{x}=v_{y}=p$ for all $x,y$, then the unique equilibrium price is $p$, so
that all consumers and producers are indifferent in equilibrium. For any
choice of coefficients $\alpha_{x}\left(  z\right)  $ and $\beta_{y}\left(
z\right)  $ such that:%
\[
0\leq\alpha_{x}\left(  z\right)  \leq1,\ \ 0\leq\beta_{y}\left(  z\right)
\leq1,\ \ \sum_{x}\alpha_{x}\left(  z\right)  =\sum_{y}\beta_{y}\left(
z\right)
\]
$\left(  p,\alpha,\beta\right)  $ is an equilibrium allocation.

\subsection{Example 2: more on uniqueness}

We give an example to clarify the uniqueness statement in Theorem \ref{12}.
There are three goods, $z_{1},z_{2,}z_{3}$, two consumers $x_{1},x_{2,}$ three
producers $y_{1},y_{2,}y_{3}$. The utility functions are:%
\begin{align*}
u\left(  x_{1},z_{1}\right)   &  =2,\ u\left(  x_{1},z_{2}\right)
=1,\ u\left(  x_{1},z_{3}\right)  =0.1\\
u\left(  x_{2},z_{1}\right)   &  =3,\ u\left(  x_{2},z_{2}\right)
=2,\ u\left(  x2,z_{3}\right)  =0.1
\end{align*}
and the cost functions are:%
\begin{align*}
v\left(  y_{1},z_{1}\right)   &  =0,\ v\left(  y_{1},z_{2}\right)
=5,\ v\left(  y_{1},z_{3}\right)  =5\\
v\left(  y_{2},z_{1}\right)   &  =5,\ v\left(  y_{2},z_{2}\right)
=0,\ v\left(  y_{2},z_{3}\right)  =5\\
v\left(  y_{3},z_{1}\right)   &  =5,\ v\left(  y_{1},z_{2}\right)
=5,\ v\left(  y_{1},z_{3}\right)  =0
\end{align*}

It is easy to check that there are two equilibria:

\begin{enumerate}
\item $y_{1}$ produces $z_{1}$, $y_{2}$ produces $z_{2}$, $y_{3}$ produces
nothing; $x_{1}$ chooses $z_{1}$, $x_{2}$ chooses $z_{2}$; prices are
$p\left(  z_{1}\right)  =1,$ $p\left(  z_{2}\right)  =0,$ $p\left(
z_{3}\right)  =0 $

\item $y_{1}$ produces $z_{1}$, $y_{2}$ produces $z_{2}$, $y_{3}$ produces
nothing; $x_{1}$ chooses $z_{2}$, $x_{2}$ chooses $z_{1}$; prices are
$p\left(  z_{1}\right)  =1.9,$ $p\left(  z_{2}\right)  =0.9,$ $p\left(
z_{3}\right)  =0$
\end{enumerate}

The demand set of $x_{1}$ is $\left\{  z_{1},z_{2}\right\}  :=D_{1}\left(
x_{1}\right)  $ in the first equilibrium and $\left\{  z_{1},z_{2}%
,z_{3}\right\}  :=D_{2}\left(  x_{1}\right)  $ in the second. The demand
\emph{distribution}, on the other hand, is $P_{x_{1}}^{1}\left(  z\right)
=\delta_{z_{1}}$ (Dirac mass at $z_{1}$) in the first equilibrium (simply
expressing the fact that $x_{1}$ chooses $z_{1}$ and nothing else in her
demand set) and $P_{x_{1}}^{2}\left(  z\right)  =\delta_{z_{2}}$ in the
second. Theorem \ref{12} then states that $\delta_{z_{2}}\left[  D_{1}\left(
x_{1}\right)  \right]  =\delta_{z_{1}}\left[  D_{1}\left(  x_{1}\right)
\right]  =1$, which simply expresses the fact that both $z_{1}$ and $z_{2}$
belong to $D_{1}\left(  x_{1}\right)  $.

Note for instance that the social utility is the same for both equilibria,
namely $4$:

\begin{enumerate}
\item In the first one:%
\[
u\left(  x_{1},z_{1}\right)  -v\left(  y_{1},z_{1}\right)  +u\left(
x_{2},z_{2}\right)  -v\left(  y_{2},z_{2}\right)  =2-0+2-0=4
\]

\item In the second one:%
\[
u\left(  x_{1},z_{2}\right)  -v\left(  y_{2},z_{2}\right)  +u\left(
x_{2},z_{1}\right)  -v\left(  y_{1},z_{1}\right)  =1-0+3-0=4
\]

\end{enumerate}

This is a general fact: the social utility is \emph{the same at all
equilibria}. Indeed, equilibrium prices are found by maximizing the right-hand
side of (\ref{par}): it may be achieved at different $p_{1}$ and $p_{2}$, but
the value of the maximum is the same.

\section{Pure equilibrium.\label{sec5}}

\subsection{Definition}

In equilibrium, consumers of type $x$ demand quality $z$ with probability
$P_{x}^{\alpha}\left(  z\right)  ,$ and suppliers of type $y$ supply quality
$z$ with probability $P_{y}^{\beta}\left(  z\right)  $. The equilibrium is
\emph{pure} if all agents of the same type who are in the market at the same
time are doing the same thing (buying or selling the same quality), so that
these probabilities are Dirac masses. Formally:

\begin{definition}
An equilibrium $\left(  p,\alpha_{X\times Z},\beta_{Y\times Z}\right)  $ is
\emph{pure} if:

\begin{itemize}
\item for $\mu$-almost every $x$, the set $D\left(  x\right)  \cap Z_{0}$
contains at most one point

\item for $\nu$-almost every $y$, the set $S\left(  y\right)  \cap Z_{0}$
contains at most one point
\end{itemize}
\end{definition}

Denote by $X_{p}$ the set of active or indifferent consumers. If $\left(
p,\alpha_{X\times Z},\beta_{Y\times Z}\right)  $ is a pure equilibrium, there
is a Borel map $d:X_{p}\rightarrow Z_{0}$ with $d\left(  x\right)  \in
D\left(  x\right)  $ such that, for $\mu$-almost every $x$, one and only one
of the following holds:

\begin{itemize}
\item either consumers of type $x$ are inactive, so that $D\left(  x\right)
=\varnothing_{d}$

\item or consumers of type $x$ are indifferent; then $D\left(  x\right)
=\varnothing_{d}\cup\left\{  d\left(  x\right)  \right\}  $

\item or consumers of type $x$ are active; then $D\left(  x\right)  =\left\{
d\left(  x\right)  \right\}  $
\end{itemize}

We can then rewrite the definition of equilibrium directly in terms of $s$ and
$d.$

\begin{definition}
A \emph{pure} \emph{equilibrium} is a triplet $\left(  p,d,s\right)  $ where:

\begin{enumerate}
\item $d$ is a Borel map from the set $X_{p}=\left\{  x\ |\ p^{\sharp}\left(
x\right)  \geq0\right\}  $ into $Z_{0}$

\item $s$ is a Borel map from the set $Y_{p}=\left\{  y\ |\ p^{\flat}\left(
y\right)  \leq0\right\}  $ into $Z_{0}$

\item For $\mu$- almost every $x$ with $p^{\sharp}\left(  x\right)  >0$, the
function $z\rightarrow u\left(  x,z\right)  -p\left(  z\right)  $ attains its
maximum at a single point $z=d\left(  x\right)  \in Z_{0}$

\item For $\nu$-almost every $y$ with $p^{\flat}\left(  y\right)  <0$, the
function $z\rightarrow p\left(  z\right)  -v\left(  y,z\right)  $ attains its
maximum at a single point $z=s\left(  y\right)  \in Z_{0}$

\item For $\mu$- almost every $x$ with $p^{\sharp}\left(  x\right)  =0$, the
function $z\rightarrow u\left(  x,z\right)  -p\left(  z\right)  $ attains its
maximum at two points, $\varnothing_{d}$ and $z=d\left(  x\right)  \in Z_{0}$

\item For $\nu$-almost every $y$ with $p^{\flat}\left(  y\right)  =0$, the
function $z\rightarrow p\left(  z\right)  -v\left(  y,z\right)  $ attains its
maximum at two points, $\varnothing_{s}$ and $z=s\left(  y\right)  \in Z_{0}$

\item The demand and supply distributions $\alpha$ and $\beta$ associated with
$d$ and $s$ have the same marginals on $Z_{0}$:%
\[
\forall A\subset Z_{0},\ \ \ \mu\left[  x\ |\ d\left(  x\right)  \in A\right]
=\nu\left[  y\ |\ s\left(  y\right)  \in A\right]
\]

\end{enumerate}
\end{definition}

For the sake of simplicity, we shall now assume that $a\left(  z\right)
<b\left(  z\right)  $ for every $z\in Z$. As a consequence, $Z_{1}=Z$.

\subsection{Uniqueness}

\begin{theorem}
\label{th8}Let $\left(  p_{1},d_{1},s_{1}\right)  $ and $\left(  p_{2}%
,d_{2},s_{2}\right)  $ be two pure equilibria. Every consumer $x$ who is
active in one equilibrium is active or indifferent in the other, and we have
$d_{1}\left(  x\right)  =d_{2}\left(  x\right)  $. Similarly, every producer
$y$ who is active in one equilibrium is active or indifferent in the other,
and $s_{1}\left(  y\right)  =s_{2}\left(  y\right)  .$
\end{theorem}

\begin{proof}
It is an immediate consequence of the uniqueness theorem for equilibrium allocations.
\end{proof}

\subsection{Existence}

\begin{theorem}
\label{thm7}Assume that the standard assumptions hold. Assume moreover that
$\mu$ and $\nu$ are absolutely continuous with respect to the Lebesgue
measure, and that the partial derivatives $D_{x}u$ and $D_{y}v$ with respect
to $z$ are injective:%
\begin{align}
\forall x  &  \in X,\;D_{x}u\left(  x,z_{1}\right)  =D_{x}u\left(
x,z_{2}\right)  \Longrightarrow z_{1}=z_{2}\label{eq11}\\
\forall y  &  \in Y,\;D_{y}v\left(  y,z_{1}\right)  =D_{y}v\left(
y,z_{2}\right)  \Longrightarrow z_{1}=z_{2} \label{eq12}%
\end{align}
Then any equilibrium is pure.
\end{theorem}

\begin{corollary}
In the above situation, there is a pure equilibrium.
\end{corollary}

\begin{proof}
We know that there is an equilibrium, by the Existence Theorem, and we know
that it has to be pure.
\end{proof}

If $X$ and $Z$ are one-dimensional intervals, condition (\ref{eq11}) is
satisfied if
\[
\frac{\partial^{2}u}{\partial x\partial z}\neq0
\]
so that condition (\ref{eq11}), or (\ref{eq12}) for that matter, is a
multi-dimensional generalization of the classical Spence-Mirrlees condition in
the economics of assymmetric information (see \cite{Carlier}). It is
satisfied, for instance, by $u\left(  x,z\right)  =\left\Vert x-z\right\Vert
^{\alpha}$, provided $\alpha\neq0$ and $\alpha\neq1$; if $\alpha<1$, one
should add the requirement that $X\cap Z\neq\varnothing$, so that $u$ is
differentiable on $X\times Z$.

\subsection{Example\label{sec6}}

\subsubsection{A case when $Z_{a}=\varnothing=Z_{b}$}

Set $X=\left[  1,2\right]  $ and $Y=\left[  2,3\right]  $. Both are endowed
with the Lebesgue measure. Set $Z_{0}=\left[  0,1\right]  $ and\label{enc}%
\begin{align*}
u\left(  x,z\right)   &  =-\frac{1}{2}z^{2}+xz,\ \ \bar{u}\left(  x\right)
=0\\
v\left(  y,z\right)   &  =\frac{1}{2}yz^{2},\ \bar{v}\left(  y\right)  =0
\end{align*}
so that suppliers are ordered on the line according to efficiency, the most
efficient ones (those with the lowest cost, near $y=2$) being on the left, and
consumers are ordered according to taste, the most avid ones (those with the
highest utility, near $x=2$) being on the right (note the order reversal).

We compute the lowest ask $a\left(  z\right)  $ and the highest bid $b\left(
z\right)  $:%
\begin{align*}
b\left(  z\right)   &  =\bar{u}^{\sharp}\left(  z\right)  =\max_{1\leq x\leq
2}\left\{  -\frac{1}{2}z^{2}+xz-0\right\}  =-\frac{1}{2}z^{2}+2z\\
a\left(  z\right)   &  =\bar{v}^{\flat}\left(  z\right)  =\min_{2\leq y\leq
3}\left\{  \frac{1}{2}yz^{2}-0\right\}  =z^{2}%
\end{align*}
Note that $b\left(  z\right)  $ is the bid price for consumer $x=2$ (the most
avid one), and $a\left(  z\right)  $ is the ask price for supplier $y=2$ (the
least efficient one). We have $a\leq b$ as expected.

Note that the generalized Spence-Mirrlees assumptions (\ref{eq11}) and
(\ref{eq12}) are satisfied:%
\begin{align*}
D_{x}u\left(  x,z\right)   &  =z\\
D_{y}v\left(  y,z\right)   &  =\frac{1}{2}z^{2}%
\end{align*}
and both are injective with respect to $z$. So Theorem \ref{thm7} applies, and
there is a pure equilibrium, with some degree of uniqueness.We shall now
compute it.

Assume for the moment that every agent is active. This is possible here since
$\mu\left(  X\right)  $ happens to be equal to $\nu\left(  Y\right)  $ (in
other words, there are as many consumers as suppliers). This means that
$Z_{a}=\varnothing=Z^{b}$, and $Z_{1}=Z_{0}$, so that we can try the reduction
method we described in the preceding section.

We start with finding the optimal matching between $X$ and $Y$. Given $x$ and
$y$, the quality $z\left(  x,y\right)  $ which maximizes the utility of the
pair $\left(  x,y\right)  $ is obtained by maximizing the expression
$-z^{2}/2+xz-yz^{2}/2$ with respect to $z$, which yields:
\begin{align*}
z\left(  x,y\right)   &  =\frac{x}{1+y}\\
w\left(  x,y\right)   &  =\frac{1}{2}\frac{x^{2}}{1+y}%
\end{align*}
where $w\left(  x,y\right)  $ is the resulting utility for the pair. We then
seek the measure-preserving map $\sigma:\left[  1,2\right]  \rightarrow\left[
2,3\right]  $ which maximizes the integral:%
\[
\int_{1}^{2}w\left(  x,\sigma\left(  x\right)  \right)  dx=\int_{1}^{2}%
\frac{x^{2}}{1+\sigma\left(  x\right)  }dx
\]

We have:%
\[
\frac{\partial^{2}w}{\partial x\partial y}=-\frac{x}{\left(  1+y\right)  ^{2}%
}<0
\]
so $w$ satisfies the Spence-Mirrlees assumption. By the general theory of
optimal transportation (see \cite{Villa}), the map $\sigma$ is uniquely
defined. We find that:%
\[
\sigma\left(  x\right)  =y=4-x
\]
either by deciding that $\sigma$ must be continous and comparing directly the
two candidates $y=4-x$ (decreasing) and $y=x+1$ (increasing), or, more
rigorously, by checking directly that $\sigma$ is the subgradient of a
$w$-convex function, which, by the general theory again, implies that it is
the minimizer. Hence the supply and demand maps $s\left(  y\right)  $ and
$d\left(  x\right)  $:%
\begin{align}
d\left(  x\right)   &  =\frac{x}{5-x}\label{d1}\\
s\left(  y\right)   &  =\frac{4-y}{1+y} \label{d2}%
\end{align}
and the set of traded qualities is $Z_{t}=\left[  \frac{1}{4},\ \frac{2}%
{3}\right]  $, which is a strict subset of $Z_{0}$: again, not all
technologically feasible quailities are traded in equilibrium. On $Z_{t}$, the
price is uniquely defined, and is found by writing the first-order condition
for optimality, $p^{\prime}\left(  z\right)  =\frac{\partial u}{\partial
z}\left(  x,z\right)  $ where $z=d\left(  x\right)  $. Inverting this map, we
get a differential equation for $p$, namely $p^{\prime}\left(  z\right)
=z+5z\left(  1+z\right)  ^{-1}$, yielding:
\begin{equation}
p\left(  z\right)  =-\frac{1}{2}z^{2}+5z-5\ln\left(  z+1\right)  +c\ \text{
for }\frac{1}{4}\leq z\leq\frac{2}{3} \label{t1}%
\end{equation}

We can now try to validate our assumption that every agent is active. Compute
the indirect utilities:%

\begin{align*}
p^{\sharp}\left(  x\right)   &  =u\left(  x,d\left(  x\right)  \right)
-p\left(  d\left(  x\right)  \right)  =x+5\left(  \ln5-\ln\left(  5-x\right)
\right)  -c\\
-p^{\flat}\left(  y\right)   &  =p\left(  s\left(  y\right)  \right)
-v\left(  y,s\left(  y\right)  \right)  =\frac{\left(  4-y\right)  \left(
6+y\right)  }{2\left(  1+y\right)  }-5\left(  \ln5-\ln\left(  1+y\right)
\right)  +c
\end{align*}

Every agent is active if and only if $p^{\sharp}\left(  x\right)  >\bar
{u}\left(  x\right)  \ $for every $x$ and $p^{\flat}\left(  y\right)
<-\bar{v}\left(  y\right)  $ for every $y$. This leads us to explicit bounds
for $c$:%

\begin{equation}
-0.00928=-\frac{9}{8}+5\ln\frac{5}{4}\leq c\leq1+5\ln\frac{5}{4}=2.1157
\label{price}%
\end{equation}

For any $c$ in that interval, the function $p\left(  z\right)  $ given by
formula (\ref{t1}) is the restriction to $Z_{t}=\left[  \frac{1}{4},\ \frac
{2}{3}\right]  $ of an equilibrium price, the equilibrium supply and demand
being given by (\ref{d2}) and (\ref{d1}).

We now have to extend $p_{t}$ to $Z_{0}=\left[  0,1\right]  $ in such a way
that the qualities $z\in\left[  0,\ \frac{1}{4}\right]  \cup\left[  \frac
{2}{3},\ 1\right]  $ are not traded. For $z=\frac{1}{4}$, the least efficient
supplier $y=3$ provides the least avid consumer$\ x=1$, and the price of
qualities $z\leq\frac{1}{4}$ must be such that each of them prefers staying at
$\frac{1}{4}$. This yields the inequalities:%
\begin{align*}
p\left(  z\right)  -v\left(  3,z\right)   &  \leq p\left(  \frac{1}{4}\right)
-v\left(  3,\frac{1}{4}\right) \\
u\left(  1,z\right)  -p\left(  z\right)   &  \leq u\left(  1,\frac{1}%
{4}\right)  -p\left(  \frac{1}{4}\right)
\end{align*}
and hence:%
\begin{equation}
-\frac{1}{2}z^{2}+z+1-5\ln\frac{5}{4}+c\leq p\left(  z\right)  \leq\frac{3}%
{2}z^{2}+\frac{9}{8}-5\ln\frac{5}{4}+c\text{ \ for }0\leq z\leq\frac{1}{4}
\label{t21}%
\end{equation}

Similarly, for $z\geq\frac{2}{3}$, we get the inequalities:%
\begin{equation}
-\frac{1}{2}z^{2}+2z+2-5\ln\frac{5}{3}+c\leq p\left(  z\right)  \leq
z^{2}+\frac{8}{3}-5\ln\frac{5}{3}+c\text{ \ for }\frac{2}{3}\leq z\leq1
\label{t22}%
\end{equation}

In summary, given any $c$ satisfying (\ref{price}), any function $p\left(
z\right)  $ satisfying (\ref{t1}), (\ref{t21}), and (\ref{t22}) is an
equilibrium price. is an equilibrium price. By Theorem \ref{th8}, $s$ and $d$
are uniquely determined, in the sense that any pure equilibrium such that all
agents are active will have the same supply and demand. This implies that the
pure equilibria we have just found are the only ones for which $Z_{0}=\bar{Z}$.

\subsubsection{A case when $Z_{a}$ is non-empty}

Let us now increase the number of consumers: say $Y=\left[  2,\ 3\right]  $ is
unchanged, while $X=\left[  h,\ 2\right]  $ with $0<h<1$. Both intervals are
endowed with the Lebesgue measure. In equilibrium, if all suppliers are
active, then consumers in the range $\left[  h,\ 1\right]  $ must be priced
out of the market. This is done by fixing $c$ in formula (\ref{price}) to its
highest possible value, namely $1+5\ln\frac{5}{4}$:%

\begin{equation}
p\left(  z\right)  =-\frac{1}{2}z^{2}+5z-5\ln\frac{4\left(  z+1\right)  }%
{5}+1\text{ for }\frac{1}{4}\leq z\leq\frac{2}{3} \label{prac}%
\end{equation}

Then consumer $x=1$ makes precisely his/her reservation utility, which means
that he/she is indifferent.

Recall that $d\left(  1\right)  =\frac{1}{4}=s\left(  3\right)  $. For
$0<z<\frac{1}{4}$, consider the bid price for quality $z$ by consumer $x=1$:%
\[
b\left(  1,z\right)  =-\frac{1}{2}z^{2}+z
\]
Consumers of type $x<1$ will have a lower bid price. Choose a continuous
function $p$ such that:%
\begin{equation}
-\frac{1}{2}z^{2}+z<p\left(  z\right)  <p\left(  \frac{1}{4}\right)  -v\left(
3,\frac{1}{4}\right)  +v\left(  3,z\right)  =\frac{3}{2}z^{2}+\frac{17}%
{8}\text{ \ for }0\leq z\leq\frac{1}{4} \label{proc}%
\end{equation}
The left inequality ensures that consumers or type $x<1$ are not bidders for
quality $z$, so they just buy quality $0$ at price $0$, that is, they revert
to their reservation utility. The right inequality ensures that the least
efficient producer will not become interested in producing quality $z,$ so
that the more efficient ones will not either.

Any function $p\left(  z\right)  $ satisfying (\ref{prac}), (\ref{proc})\ and
(\ref{t22})\ (with $1+5\ln\frac{5}{4}$) is an equilibrium price. Note that for
all consumers $x\in\lbrack h,\ 1[$ demand $\ $is uniquely defined: $d\left(
x\right)  =0$.

\section{Open problems.}

In this paper, we have assumed that the good is indivisible, and that
consumers and producers are limited to buying and selling one unit. That
assumption can be relaxed. Indeed, our results carry through if we assume that
suppliers, for instance, are restricted to producing one quality, but have the
choice of the quantity they produce, their profit then being $np-v\left(
y,z,n\right)  $, where $z$ is the quality produced, $n$ the quantity, $p$ the
price, and $y$ the type of the supplier.

As we mentioned in the beginning, the main limitation of our model is the
assumption that utilities are separable. A truly general model would introduce
a quantity good beside the quality good, and consumers of type $x$ would solve
the problem:%
\[
\max\left\{  u\left(  x,z,t\right)  \ |\ p\left(  z\right)  +\pi t\leq
w\right\}
\]
where $t$ is the quantity of the second good, and $\pi$ its (linear) price.
Our methods do not readily apply to this situation, and we plan to investigate
it further.

Finally, we wish to stress that although we have what appears as a complete
equilibrium theory for multidimensional hedonic models, the numerical aspects
are far from being as well understood. The method we used in the example is
strictly one-dimensional, and there is no easy way to extend it to the
multidimensional case. The obvious way to proceed is to follow the theoretical
argument, and try to minimize the integral $I\left(  p\right)  $ in
(\ref{uni}), but we have made no progress in that direction. It certainly is a
good topic for future research. So will all the econometric aspects
(characterization and identification). This investigation has been started in
\cite{Ekel}, but is far from being complete.

\appendix

\section{Fundamentals of $u$-convex analysis.\label{sec1}}

In this section, we basically follow Carlier \cite{Carlier}.

\subsection{$u$-convex functions.}

We will be dealing with function taking values in $\mathbb{R\cup}\left\{
+\infty\right\}  $.

A function $f:X\rightarrow\mathbb{R\cup}\left\{  +\infty\right\}  $ will be
called $u$\emph{-convex} iff there exists a non-empty subset $A\subset
Z\times\mathbb{R}$ such that:
\begin{equation}
\forall x\in X,\;\;f\left(  x\right)  =\sup_{\left(  z,\alpha\right)  \in
A}\left\{  u\left(  x,z\right)  +a\right\}  \label{u-con}%
\end{equation}

A function $p:Z\rightarrow\mathbb{R\cup}\left\{  +\infty\right\}  $ will be
called $u$\emph{-convex} iff there exists a non-empty subset $B\subset
X\times\mathbb{R}$ such that:
\begin{equation}
p\left(  z\right)  =\sup_{\left(  x,b\right)  \in B}\left\{  u\left(
x,z\right)  +b\right\}  \label{u-conv*}%
\end{equation}

\subsection{Subconjugates}

Let $f:X\rightarrow\mathbb{R\cup}\left\{  +\infty\right\}  $, not identically
$\left\{  +\infty\right\}  $, be given. We define its \emph{subconjugate}
$f^{\sharp}:Z\rightarrow\mathbb{R\cup}\left\{  +\infty\right\}  $ by:
\begin{equation}
f^{\sharp}\left(  z\right)  =\sup_{x}\left\{  u\left(  x,z\right)  -f\left(
x\right)  \right\}  \label{Fen}%
\end{equation}

It follows from the definitions that $f^{\sharp}$ is a $u$-convex\emph{\ }%
function on $Z$ (it might be identically $\left\{  +\infty\right\}  $).

Let $p:Z\rightarrow\mathbb{R\cup}\left\{  +\infty\right\}  $, not identically
$\left\{  +\infty\right\}  $, be given. We define its \emph{subconjugate}
$p^{\sharp}:X\rightarrow\mathbb{R\cup}\left\{  +\infty\right\}  $ by:
\begin{equation}
p^{\sharp}\left(  x\right)  =\sup_{z}\left\{  u\left(  x,z\right)  -p\left(
z\right)  \right\}  \label{Fen*}%
\end{equation}

It follows from the definitions that $p^{\sharp}$ is a $u$-convex\emph{\ }%
function on $X$\ (it might be identically $\left\{  +\infty\right\}  $).

\begin{example}
Set $f\left(  x\right)  =u\left(  x,\bar{z}\right)  +a$. Then
\[
f^{\sharp}\left(  \bar{z}\right)  =\sup_{x}\left\{  u\left(  x,\bar{z}\right)
-u\left(  x,\bar{z}\right)  -a\right\}  =-a
\]

\end{example}

Conjugation reverses ordering: if $f_{1}\leq f_{2}$, then $f_{1}^{\sharp}\geq
f_{2}^{\sharp}$, and if $p_{1}\leq p_{2},$ then $p_{1}^{\sharp}\geq
p_{2}^{\sharp}$. As a consequence, if $f$ is $u$-convex, not identically
$\left\{  +\infty\right\}  $, then $f^{\sharp}$ is $u$-convex, not identically
$\left\{  +\infty\right\}  $,. Indeed, since $f$ is $u$-convex, we have
$f\left(  x\right)  \geq u\left(  x,z\right)  +a\;$for some $\left(
z,a\right)  $, and then $f^{\sharp}\left(  z\right)  \leq-a<\infty.$

\begin{proposition}
[the Fenchel inequality]For any functions $f:X\rightarrow\mathbb{R\cup
}\left\{  +\infty\right\}  $ and $p:Z\rightarrow\mathbb{R\cup}\left\{
+\infty\right\}  $, not identically $\left\{  +\infty\right\}  $, we have:
\begin{align*}
\forall\left(  x,z\right)  ,\;\;f\left(  x\right)  +f^{\sharp}\left(
z\right)   &  \geq u\left(  x,z\right)  \;\;\\
\forall\left(  x,z\right)  \;\;p\left(  z\right)  +p^{\sharp}\left(  x\right)
&  \geq u\left(  x,z\right)  \;\;
\end{align*}

\end{proposition}

\subsection{Subgradients}

Let $f:X\rightarrow\mathbb{R\cup}\left\{  +\infty\right\}  $ be given, not
identically $\left\{  +\infty\right\}  $. Take some point $x\in X$. We shall
say that a point $z\in Z$ is a \emph{subgradient} of $f$ at $x$ if the points
$x$ and $z$ achieve equality in the Fenchel inequality:
\begin{equation}
f\left(  x\right)  +f^{\sharp}\left(  z\right)  =u\left(  x,z\right)
\label{subd}%
\end{equation}

The set of subgradients of $f$ at $x$ will be called the
\emph{subdifferential} of $f$ at $x$ and denoted by $\partial f\left(
x\right)  $. Specifically:

\begin{definition}
$\partial f\left(  x\right)  =\arg\max_{z}\left\{  u\left(  x,z\right)
-f^{\sharp}\left(  z\right)  \ \right\}  $
\end{definition}

Similarly, let $p:Z\rightarrow\mathbb{R\cup}\left\{  +\infty\right\}  $ be
given, not identically $\left\{  +\infty\right\}  $. Take some point $z\in Z$.
We shall say that a point $x\in X$ is a \emph{subgradient} of $p$ at $z$ if:
\begin{equation}
p^{\sharp}\left(  x\right)  +p\left(  z\right)  =u\left(  x,z\right)
\label{subdd}%
\end{equation}
The set of subgradients of $p$ at $z$ will be called the
$\emph{subdifferential}$ of $p$ at $z$ and denoted by $\partial p\left(
z\right)  $.

\begin{definition}
$\partial p\left(  z\right)  =\arg\max_{x}\left\{  u\left(  x,z\right)
-p^{\sharp}\left(  x\right)  \right\}  $
\end{definition}

\begin{proposition}
\label{prop1}The following are equivalent:

\begin{enumerate}
\item $z\in\partial f\left(  x\right)  $

\item $\forall x^{\prime},\;\;f\left(  x^{\prime}\right)  \geq f\left(
x\right)  +u\left(  x^{\prime},z\right)  -u\left(  x,z\right)  \;\;$
\end{enumerate}

If equality holds for some $x^{\prime}$, then $z\in\partial f\left(
x^{\prime}\right)  $ as well.
\end{proposition}

\begin{proof}
We begin with proving that the first condition implies the second one. Assume
$z\in\partial f\left(  x\right)  $. Then, by (\ref{subd}) and the Fenchel
inequality, we have:
\[
f\left(  x^{\prime}\right)  \geq u\left(  x^{\prime},z\right)  -f^{\sharp
}\left(  z\right)  =u\left(  x^{\prime},z\right)  -\left[  u\left(
x,z\right)  -f\left(  x\right)  \right]
\]

We then prove that the second condition implies the first one. Using the
inequality, we have:
\begin{align*}
f^{\sharp}\left(  z\right)   &  =\sup_{x^{\prime}}\left\{  u\left(  x^{\prime
},z\right)  -f\left(  x^{\prime}\right)  \right\} \\
&  \leq\sup_{x^{\prime}}\left\{  u\left(  x^{\prime},z\right)  -f\left(
x\right)  -u\left(  x^{\prime},z\right)  +u\left(  x,z\right)  \right\} \\
&  =u\left(  x,z\right)  -f\left(  x\right)
\end{align*}
so $f\left(  x\right)  +f^{\sharp}\left(  z\right)  \leq u\left(  x,z\right)
$. We have the converse by the Fenchel inequality, so equality holds.

Finally, if equality holds for some $x^{\prime}$ in condition (2), then
$\;f\left(  x^{\prime}\right)  -u\left(  x^{\prime},z\right)  =f\left(
x\right)  -u\left(  x,z\right)  $, so that:
\begin{align*}
\forall x^{\prime\prime},\;\;f\left(  x^{\prime\prime}\right)   &  \geq
f\left(  x\right)  -u\left(  x,z\right)  +u\left(  x^{\prime\prime},z\right)
\\
&  =f\left(  x^{\prime}\right)  -u\left(  x^{\prime},z\right)  +u\left(
x^{\prime\prime},z\right)
\end{align*}
which implies that $z\in\partial f\left(  x^{\prime}\right)  $.
\end{proof}

There is a similar result for functions $p:Z\rightarrow\mathbb{R\cup}\left\{
+\infty\right\}  $, not identically $\left\{  +\infty\right\}  $: we have
$x\in\partial p\left(  z\right)  $ if and only if
\begin{equation}
\forall\left(  x^{\prime},\bar{Z}\right)  ,\;\;p\left(  \bar{Z}\right)  \geq
p\left(  z\right)  +u\left(  x,\bar{Z}\right)  -u\left(  x,z\right)  \;\;
\label{eq5}%
\end{equation}

\subsection{Biconjugates}

It follows from the Fenchel inequality that, if $p:Z\rightarrow\mathbb{R\cup
}\left\{  +\infty\right\}  $ is not identically $\left\{  +\infty\right\}  $:%

\begin{equation}
p^{\sharp\sharp}\left(  z\right)  =\sup_{x}\left\{  u\left(  x,z\right)
-p^{\sharp}\left(  x\right)  \right\}  \leq p\left(  z\right)  \label{eq7}%
\end{equation}

\begin{example}
Set $p\left(  z\right)  =u\left(  \bar{x},z\right)  +b$. Then
\begin{align*}
p^{\sharp\sharp}\left(  z\right)   &  =\sup_{x}\left\{  u\left(  x,z\right)
-p^{\sharp}\left(  x\right)  \right\} \\
&  \geq u\left(  \bar{x},z\right)  -p^{\sharp}\left(  \bar{x}\right) \\
&  =u\left(  \bar{x},z\right)  +b=p\left(  z\right)
\end{align*}

\end{example}

This example generalizes to all $u$-convex functions. Denote by $\mathbb{C}%
_{u}\left(  Z\right)  $ the set of all $u$-convex functions on $Z$.

\begin{proposition}
\label{prop2}For every function $p:Z\rightarrow\mathbb{R\cup}\left\{
+\infty\right\}  $, not identically $\left\{  +\infty\right\}  $, we have
\[
p^{\sharp\sharp}\left(  z\right)  =\sup_{\varphi}\left\{  \varphi\left(
z\right)  \;\left|  \;\varphi\leq p,\;\varphi\in\mathbb{C}_{u}\left(
Z\right)  \right.  \right\}
\]

\end{proposition}

\begin{proof}
Denote by $\bar{p}\left(  z\right)  $ the right-hand side of the above
formula. We want to show that $p^{\sharp\sharp}\left(  z\right)  =\bar
{p}\left(  z\right)  $

Since $p^{\sharp\sharp}\leq p$ and $p^{\sharp\sharp}$ is $u$-convex, we must
have$\;p^{\sharp\sharp}\leq\bar{p}$.

On the other hand, $\bar{p}$ is $u$-convex because it is a supremum of
$u$-convex functions. So there must be some $B\subset X\times\mathbb{R}$ such
that:
\[
\bar{p}\left(  z\right)  =\sup_{\left(  x,b\right)  \in B}\left\{  u\left(
x,z\right)  +b\right\}
\]
Let $\left(  x,b\right)  $ $\in B$. Since $\bar{p}\leq p$, we have $u\left(
x,z\right)  +b\leq\bar{p}\left(  z\right)  \leq p\left(  z\right)  $. Taking
biconjugates, as in the preceding example, we get $\ u\left(  x,z\right)
+b\leq p^{\sharp\sharp}\left(  z\right)  $. Taking the supremums over $\left(
x,b\right)  $ $\in B$, we get the desired result.
\end{proof}

\begin{corollary}
\label{cor2}Let $p:Z\rightarrow\mathbb{R\cup}\left\{  +\infty\right\}  $ be a
$u$-convex function, not identically $\left\{  +\infty\right\}  $. Then
$p=p^{\sharp\sharp}$, and the following are equivalent:

\begin{enumerate}
\item $x\in\partial p\left(  z\right)  $

\item $p\left(  z\right)  +p^{\sharp}\left(  x\right)  =u\left(  x,z\right)  $

\item $z\in\partial p^{\sharp}\left(  x\right)  $
\end{enumerate}
\end{corollary}

\begin{proof}
We have $p^{\sharp\sharp}\leq p$ always by relation (\ref{eq7}). Since $p$ is
$u$-convex, we have:
\[
p\left(  z\right)  =\sup_{\left(  x,b\right)  \in B}\left\{  u\left(
x,z\right)  +b\right\}
\]
for some $B\subset X\times\mathbb{R}$. By proposition \ref{prop2}, we have:
\[
\sup_{\left(  x,b\right)  \in B}\left\{  u\left(  x,z\right)  +b\right\}  \leq
p^{\sharp\sharp}\left(  z\right)
\]
and so we must have $p=p^{\sharp\sharp}$. Taking this relation into account,
as well as the definition of the subgradient, we see that condition (2) is
equivalent both to (1) and to (3)
\end{proof}

\begin{definition}
We shall say that a function $p:Z\rightarrow\mathbb{R\cup}\left\{
+\infty\right\}  $ is $u$-adapted if it is not identically $\left\{
+\infty\right\}  $ and there is some $\left(  x,b\right)  \in X\times R$ such
that:
\[
\forall z\in Z,\;\;p\left(  z\right)  \geq u\left(  x,z\right)  +b
\]

\end{definition}

It follows from the above that if $p$ is $u$-adapted, then so are $p^{\sharp}%
$, $p^{\sharp\sharp}$ and all further subconjugates. Note that a $u$-convex
function which is not identically $\left\{  +\infty\right\}  $ is $u$-adapted.

\begin{corollary}
\label{cor3}Let $p:Z\rightarrow\mathbb{R\cup}\left\{  +\infty\right\}  $ be
$u$-adapted. Then :
\[
p^{\sharp\sharp\sharp}=p^{\sharp}%
\]

\end{corollary}

\begin{proof}
If $p$ is $u$-adapted, then $p^{\sharp}$ is $u$-convex and not identically
$\left\{  +\infty\right\}  $. The result then follows from corollary
\ref{cor2}.
\end{proof}

\subsection{Smoothness}

Since $u$ is continuous and $X\times Z$ is compact, the family
\[
\left\{  u\left(  x,\cdot\right)  \;\left\vert \;x\in X\right.  \right\}
\]
is uniformly equicontinuous on $Z$. It follows from definition \ref{u-con}
that all $u$-convex functions on $Z$ are continuous (in particular, they are
finite everywhere)..

Denote by $k$ the upper bound of $\left\Vert D_{x}u\left(  x,z\right)
\right\Vert $ for $\left(  x,z\right)  \in X\times Z$. Since $D_{x}u$ is
continuous and $X\times Z$ is compact, we have $k<\infty$, and the functions
$x\rightarrow u\left(  x,z\right)  $ are all $k$-Lipschitzian on $X$. Again,
it follows from the definition \ref{u-con} that all $u$-convex functions on
$X$ are $k$-Lipschitz (in particular, they are finite everywhere). By
Rademacher' theorem, they are differentiable almost everywhere with respect to
the Lebesgue measure.

Let $f$ $:X\rightarrow R$ be convex. Since $f=f^{\sharp\sharp}$, we have:
\[
f\left(  x\right)  =\sup_{z}\left\{  u\left(  x,z\right)  -f^{\sharp}\left(
z\right)  \right\}
\]
Since $f^{\sharp}$ is $u$-convex, it is continuous, and the supremum is
achieved on the right-hand side, at some point $z\in\partial f\left(
x\right)  $. This means that all $u$-convex functions on $X$ are
subdifferentiable everywhere on $X$.

The following result will also be useful:

\begin{proposition}
\label{prop7} Let $p:Z\rightarrow R$ be $u$-adapted, and let $x\in X$ be
given. Then there is some point $z\in\partial p^{\sharp}\left(  x\right)  $
such that $p\left(  z\right)  =p^{\sharp\sharp}\left(  z\right)  $.
\end{proposition}

\begin{proof}
Assume otherwise, so that for every $z\in\partial p^{\sharp}\left(  x\right)
$ we have $p^{\sharp\sharp}\left(  z\right)  <p\left(  z\right)  $. For every
$z\in\partial p^{\sharp}\left(  x\right)  $, we have $x\in\partial
p^{\sharp\sharp}\left(  z\right)  $, so that, by proposition \ref{prop1}, we
have
\[
p^{\sharp\sharp}\left(  z^{\prime}\right)  \geq u\left(  x,z^{\prime}\right)
-u\left(  x,z\right)  +p^{\sharp\sharp}\left(  z\right)
\]
for all $z^{\prime}\in Z$, the inequality being strict if $z^{\prime}\notin$
$\partial p^{\sharp}\left(  x\right)  .$\ Set $\varphi_{z}\left(  z^{\prime
}\right)  =u\left(  x,z^{\prime}\right)  -u\left(  x,z\right)  +p^{\sharp
\sharp}\left(  z\right)  $. We have:
\begin{align*}
z^{\prime}  &  \notin\partial p^{\sharp}\left(  x\right)  \Longrightarrow
\varphi_{z}\left(  z^{\prime}\right)  <p^{\sharp\sharp}\left(  z^{\prime
}\right)  \leq p\left(  z^{\prime}\right) \\
z^{\prime}  &  \in\partial p^{\sharp}\left(  x\right)  \Longrightarrow
\varphi_{z}\left(  z^{\prime}\right)  \leq p^{\sharp\sharp}\left(  z^{\prime
}\right)  <p\left(  z^{\prime}\right)
\end{align*}
so that $\varphi_{z}\left(  z^{\prime}\right)  <p\left(  z^{\prime}\right)  $
for all $\left(  z,z^{\prime}\right)  $. Since $Z$ is compact, there is some
$\varepsilon>0$ such that $\varphi_{z}\left(  z^{\prime}\right)
+\varepsilon\leq p\left(  z^{\prime}\right)  $ for all $\left(  z,z^{\prime
}\right)  $. Taking the subconjugate with respect to $z^{\prime}$, we get:
\begin{align*}
p^{\sharp}\left(  x\right)   &  \leq\sup_{z^{\prime}}\left\{  u\left(
x,z^{\prime}\right)  -\varphi_{z}\left(  z^{\prime}\right)  \right\}
-\varepsilon\\
&  =\sup_{z^{\prime}}\left\{  u\left(  x,z^{\prime}\right)  -u\left(
x,z^{\prime}\right)  +u\left(  x,z\right)  -p^{\sharp\sharp}\left(  z\right)
\right\}  -\varepsilon\\
&  =u\left(  x,z\right)  -p^{\sharp\sharp}\left(  z\right)  -\varepsilon
=p^{\sharp}\left(  x\right)  -\varepsilon
\end{align*}
which is a contradiction. The result follows
\end{proof}

\begin{corollary}
\label{cor7}If $\partial p^{\sharp}\left(  x\right)  =\left\{  z\right\}  $ is
a singleton, then:
\begin{equation}
p\left(  z\right)  =p^{\sharp\sharp}\left(  z\right)  \label{eq25}%
\end{equation}
and:
\begin{equation}
p^{\sharp}\left(  x\right)  =u\left(  x,z\right)  -p\left(  z\right)
\label{eq26}%
\end{equation}

\end{corollary}

\begin{proof}
Just apply the preceding proposition, bearing in mind that $\partial
p^{\sharp}\left(  x\right)  $ contains only one point, namely $\nabla
_{u}p^{\sharp}\left(  x\right)  $. This yields equation (\ref{eq25}) Equation
(\ref{eq26}) follows from the definition of the subgradient and equation
(\ref{eq25}).
\end{proof}

\subsection{$v$-concave functions.}

Let us now consider the duality between $Y$ and $Z$. Given $v:Y\times
Z\rightarrow R$, we say that a map $g:Y\rightarrow\mathbb{R\cup}\left\{
-\infty\right\}  $ is $v$\emph{-concave} iff there exists a non-empty subset
$A\subset Z\times\mathbb{R}$ such that:
\begin{equation}
\forall y\in Y,\;\;g\left(  y\right)  =\inf_{\left(  z,a\right)  \in
A}\left\{  v\left(  y,z\right)  +a\right\}
\end{equation}
and a function $p:Z\rightarrow\mathbb{R\cup}\left\{  -\infty\right\}  $ will
be called $v$\emph{-concave} iff there exists a non-empty subset $B\subset
X\times\mathbb{R}$ such that:
\begin{equation}
p\left(  z\right)  =\inf_{\left(  x,b\right)  \in B}\left\{  v\left(
y,z\right)  +b\right\}
\end{equation}

All the results on $u$-convex functions carry over to $v$-concave functions,
with obvious modifications. The \emph{superconjugate} of a function
$g:Y\rightarrow\mathbb{R\cup}\left\{  -\infty\right\}  $, not identically
$\left\{  -\infty\right\}  $, is defined by:
\begin{equation}
g^{\flat}\left(  z\right)  =\inf_{y}\left\{  v\left(  y,z\right)  -g\left(
y\right)  \right\}
\end{equation}
and the \emph{superconjugate} of a function $p:Z\rightarrow\mathbb{R\cup
}\left\{  -\infty\right\}  $, not identically $\left\{  -\infty\right\}  $, is
given by:
\begin{equation}
p^{\flat}\left(  y\right)  =\inf_{z}\left\{  v\left(  y,z\right)  -p\left(
z\right)  \right\}
\end{equation}

The superdifferential $\partial p^{\flat}$ is defined by:
\[
\partial p^{\flat}\left(  y\right)  =\arg\min_{z}\left\{  v\left(  y,z\right)
-p\left(  z\right)  \right\}
\]
and we have the Fenchel inequality:
\[
p\left(  z\right)  +p^{\flat}\left(  y\right)  \leq v\left(  y,z\right)
\;\;\forall\left(  y,z\right)
\]
with equality iff $z\in\partial p^{\flat}\left(  y\right)  $. Note finally
that $p^{\flat\flat}\geq p$, with equality if $p$ is $v$-concave

\section{Some notations and definitions.\label{sec23}}

\subsection{Radon measures and probabilities.}

With a locally compact set $\Omega$ (such as an open subset of the compact set
$Z$) we will associate the following sets of functions and measures on
$\Omega$:

\begin{itemize}
\item $\mathcal{K}\left(  \Omega\right)  $, the space of continous functions
on $\Omega$ with compact support

\item $\mathcal{C}^{b}\left(  \Omega\right)  \,$, the space of bounded
continous functions on $\Omega$

\item $\mathcal{C}_{+}\left(  \Omega\right)  $, the cone of non-negative functions

\item $\mathcal{M}\left(  \Omega\right)  $, the space of measures on $\Omega$

\item $\mathcal{M}_{+}\left(  \Omega\right)  \subset\mathcal{M}\left(
\Omega\right)  $, the cone of positive measures

\item $\mathcal{M}_{b}\left(  \Omega\right)  \subset\mathcal{M}\left(
\Omega\right)  $, the cone of finite measures

\item $\mathcal{M}_{+}^{b}\left(  \Omega\right)  =\mathcal{M}_{b}\left(
\Omega\right)  \cap\mathcal{M}_{+}\left(  \Omega\right)  $, the cone of
positive finite measures

\item $\mathcal{P}\left(  \Omega\right)  \subset\mathcal{M}_{+}^{b}\left(
\Omega\right)  $ the set of probabilities on $\Omega$
\end{itemize}

The space $\mathcal{K}\left(  \Omega\right)  $ will be endowed with the
topology of uniform convergence on compact subsets of $\Omega$, and the space
$\mathcal{C}^{b}\left(  \Omega\right)  $ with the uniform norm. Then
$\mathcal{C}^{b}\left(  \Omega\right)  $ is a Banach space, but $\mathcal{K}%
\left(  \Omega\right)  $ is not, unless $\Omega$ is compact, in which case all
continuous functions on $\Omega$ are bounded, and we have $\mathcal{C}\left(
\Omega\right)  =K\left(  \Omega\right)  =\mathcal{C}^{b}\left(  \Omega\right)
$. When $\Omega$ is finite and has $d$ elements, all these spaces coincide
with $R^{d}$.

We take measures in the sense of Radon, that is, $\mathcal{M}\left(
\Omega\right)  $ is defined to be the dual of $\mathcal{K}\left(
\Omega\right)  $ and $\mathcal{M}_{b}\left(  \Omega\right)  $ is defined to be
the dual of $\mathcal{C}^{b}\left(  \Omega\right)  $. So $\mathcal{M}%
_{b}\left(  \Omega\right)  $ is a Banach space, but $\mathcal{M}\left(
\Omega\right)  $ is not, unless $\Omega$ is compact, in which case
$\mathcal{M}\left(  \Omega\right)  =\mathcal{M}_{b}\left(  \Omega\right)  $,
that is, all Radon measures on $\Omega$ are finite. For $\gamma\in
\mathcal{M}\left(  \Omega\right)  $ and $\varphi\in\mathcal{K}\left(
\Omega\right)  $,we write indifferently $<\gamma,\varphi>$ or $\int_{Z}\varphi
d\gamma$.

A \emph{probability} $\gamma\in\mathcal{P}\left(  Z\right)  $ is defined as a
non-negative bounded measure such that $<\gamma,1>=1$. The set $\mathcal{P}%
\left(  Z\right)  $ is convex, and is compact in the weak* topology:
$\gamma_{n}\rightarrow\gamma$ if $<\gamma_{n},\varphi>\rightarrow
<\gamma,\varphi>$ for every $\varphi\in\mathcal{C}^{b}\left(  \Omega\right)  $.

We say that\ a measure $\gamma$ is \emph{carried by }$K$ if $<\gamma
,\varphi>=0$ for all $\varphi\in\mathcal{K}\left(  \Omega\right)  $ which
vanish on $K$. If $\gamma$ is carried by a subset $K$, it is also carried by
its closure. The \emph{support} of a measure $\gamma$, denoted by
$\mathrm{Supp}\left(  \gamma\right)  $, is the smallest closed set $K$ such
that $\gamma$ is carried by $K$.

\subsection{Conditional probabilities and marginals.}

Given a positive measure $\alpha_{X\times Z}\in\mathcal{M}_{+}\left(  X\times
Z\right)  $ (which has to be finite, since $X\times Z$ is compact) we define
its marginals $\alpha_{X}\in\mathcal{M}_{+}\left(  X\right)  $ and $\alpha
_{Z}\in\mathcal{M}_{+}\left(  Z\right)  $as follows:%

\begin{align*}
\int_{X}\varphi\left(  x\right)  d\alpha_{X}  &  =\int_{X\times Z}%
\varphi\left(  x\right)  d\alpha_{X\times Z}\ \ \ \forall\varphi\in
\mathcal{K}\left(  X\right) \\
\int_{Z}\psi\left(  z\right)  d\alpha_{Z}  &  =\int_{X\times Z}\psi\left(
z\right)  d\alpha_{X\times Z}\ \ \ \forall\psi\in\mathcal{K}\left(  Z\right)
\end{align*}
and we denote the probability of the second coordinate being $z$ conditional
on the first coordinate being $x$ by $P_{x}^{\alpha}\left(  z\right)  $. The
mathematical expectation with respect to this probability will be denoted by
$E_{x}^{\alpha}$:%
\[
E_{x}^{\alpha}\left[  \psi\right]  =\int_{Z}\psi\left(  z\right)
dP_{x}^{\alpha}\left(  z\right)
\]

This conditional probability is related to the first marginal by the formula:%
\[
\int_{X\times Z}f\left(  x,z\right)  d\alpha_{X\times Y}=\int_{X}E_{x}%
^{\alpha}\left[  f\left(  x,z\right)  \right]  d\alpha_{X}\ \ \ \forall
f\in\mathcal{K}\left(  X\times Z\right)
\]

Similar considerations hold for positive measures $\beta_{Y\times Z}%
\in\mathcal{M}_{+}\left(  Y\times Z\right)  $, We have:%
\begin{align*}
\int_{Y}\varphi\left(  y\right)  d\beta_{Y}  &  =\int_{Y\times Z}%
\varphi\left(  y\right)  d\beta_{Y\times Z}\ \ \ \forall\varphi\in
\mathcal{K}\left(  Y\right) \\
\int_{Z}\psi\left(  z\right)  d\beta_{Z}  &  =\int_{X\times Z}\psi\left(
z\right)  d\beta_{Y\times Z}\ \ \ \forall\psi\in\mathcal{K}\left(  Z\right) \\
\int_{Y\times Z}g\left(  y,z\right)  d\beta_{Y\times Z}  &  =\int_{Y}%
E_{y}^{\beta}\left[  g\left(  y,z\right)  \right]  d\beta_{Y}\ \ \ \forall
g\in\mathcal{K}\left(  Y\times Z\right)
\end{align*}

\section{Proof of the Existence Theorem\label{sec20}}

\subsection{The dual problem:\ existence}

Recall that $Z=\left\{  \varnothing_{d}\right\}  \cup Z_{0}\cup\left\{
\varnothing_{s}\right\}  ,$with $Z_{1}=\left\{  z\ |\ a\left(  z\right)  \leq
b\left(  z\right)  \right\}  $ a compact non-empty subset of $Z_{0}$. Denote
by $\mathcal{A}$ the set of all admissible price systems on $Z$, that is, the
set of all continuous maps $p:Z\rightarrow R~\ $\ which satisfy:%
\[
\forall z\in Z_{1},\ \ \ a\left(  z\right)  \leq p\left(  z\right)  \leq
b\left(  z\right)
\]

$\mathcal{A}$ is a non-empty, convex and closed subset of $\mathcal{K}\left(
Z\right)  ,$ the space of all continuous functions on $Z$. Now define a map
$I:\mathcal{K}\left(  Z\right)  \rightarrow R$ by:%
\begin{equation}
I\left(  p\right)  =\int_{X}p^{\sharp}\left(  x\right)  d\mu-\int_{Y}p^{\flat
}\left(  y\right)  d\nu\label{uni}%
\end{equation}

\begin{proposition}
The map $I\ $\ is convex
\end{proposition}

\begin{proof}
Take $p_{1}$ and $p_{2}$ in $\mathcal{A}$. Take $s$ and $t$ in $\left[
0,1\right]  $ with $s+t=1$. Then:%
\begin{align*}
\left(  sp_{1}+tp_{2}\right)  ^{\sharp}\left(  x\right)   &  =\sup_{z}\left\{
u\left(  x,z\right)  -sp_{1}\left(  z\right)  -tp_{2}\left(  z\right)
\right\} \\
&  =\sup_{z}\left\{  s\left[  u\left(  x,z\right)  -p_{1}\left(  z\right)
\right]  +t\left[  u\left(  x,z\right)  -p_{2}\left(  z\right)  \right]
\right\} \\
&  \leq s\sup_{z}\left\{  u\left(  x,z\right)  -p_{1}\left(  z\right)
\right\}  +t\sup_{z}\left\{  u\left(  x,z\right)  -p_{2}\left(  z\right)
\right\} \\
&  =sp_{1}^{\sharp}\left(  x\right)  +tp_{2}^{\sharp}\left(  x\right)
\end{align*}

Similarly, we find that:%
\[
\left(  sp_{1}+tp_{2}\right)  ^{\flat}\left(  y\right)  \geq sp_{1}^{\flat
}\left(  x\right)  +tp_{2}^{\flat}\left(  x\right)
\]

Integrating, we find that $I$ is convex, as announced.
\end{proof}

Now consider the convex optimization problem:%
\begin{equation}
\inf_{p\in\mathcal{A}}I\left(  p\right)  \label{P}%
\end{equation}

\begin{proposition}
The set of solutions of problem (P) is convex.
\end{proposition}

This follows from the fact that we are minimizing a convex function on a
convex set.

We have to show that this set is non-empty. The following lemma will be useful.

\begin{lemma}
\label{lem2}Assume $p$ is admissible. Set:%
\[
p_{a}^{\sharp\sharp}\left(  z\right)  =\max\left\{  p^{\sharp\sharp}\left(
z\right)  ,a\left(  z\right)  \right\}
\]
Then $\left(  p_{a}^{\sharp\sharp}\right)  ^{\sharp}=p^{\sharp}$
\end{lemma}

\begin{proof}
We have $p^{\sharp\sharp}\leq p_{a}^{\sharp\sharp}\leq p$. Taking conjugates,
we get $\ p^{\sharp}\leq\left(  p_{a}^{\sharp\sharp}\right)  ^{\sharp}%
\leq\left(  p^{\sharp\sharp}\right)  ^{\sharp}=p^{\sharp}$.
\end{proof}

Similarly, we find that $\left(  p_{b}^{\flat\flat}\right)  ^{\flat}=p^{\flat
}$, with $p_{b}^{\flat\flat}=\min\left\{  p^{\flat\flat},b\right\}  $

\begin{proposition}
Problem (P) has a solution.
\end{proposition}

\begin{proof}
Take a minimizing sequence $p_{n}$. Since the functions $p_{n}^{\sharp}$
(resp. $p_{n}^{\flat})$, $n\in N$, are $u$-convex (resp. $v$-concave), they
are uniformly Lipschitzian (see section \ref{sec1}), and hence equicontinuous.
By Ascoli's theorem we can extract uniformly convergent subsequences (still
denoted by $p_{n}^{\sharp}$ and $p_{n}^{\flat}$) :
\begin{align*}
p_{n}^{\sharp}  &  \rightarrow f\\
p_{n}^{\flat}  &  \rightarrow g
\end{align*}
so that:
\begin{equation}
\int_{X}f\left(  x\right)  d\mu-\int_{Y}g\left(  y\right)  d\nu=\inf_{a\leq
p\leq b}\left[  \int_{X}p^{\sharp}\left(  x\right)  d\mu-\int_{Y}p^{\flat
}\left(  y\right)  d\nu\right]  \label{oq}%
\end{equation}

It is easy to see that $f$ is $u$-convex and $g$ is $v$-concave. In addition,
$p_{n}^{\sharp\sharp}\rightarrow f^{\sharp}$ and $p_{n}^{\flat\flat
}\rightarrow g^{\flat}$ everywhere (and uniformly as well, since the functions
are equicontinuous). Since $p_{n}^{\sharp\sharp}\leq p_{n}^{\flat\flat}$, we
get $f^{\sharp}$ $\leq g^{\flat}$ in the limit. Since $p_{n}\leq b$, we have
$p_{n}^{\sharp\sharp}\leq b^{\sharp\sharp}=b$, and letting $n\rightarrow
\infty$, we find that $f^{\sharp}\leq b$. Since $f^{\sharp}$ is $u$-convex, it
is continuous (and even Lipschitzian, see section \ref{sec1}). Similarly,
$g^{\flat}$ is $v$-concave, hence continuous, and satisfies $g^{\flat}\geq a$.

Now take any continuous price schedule $\bar{p}$ such that\
\begin{equation}
\left(  f^{\sharp}\right)  _{a}^{\sharp\sharp}=\max\newline\left\{  f^{\sharp
},a\right\}  \leq\bar{p}\leq\min\left\{  g^{\flat},b\right\}  =\left(
g^{\flat}\right)  _{b}^{\flat\flat} \label{qe}%
\end{equation}
for instance $\bar{p}=\frac{1}{2}\left(  \max\newline\left\{  f^{\sharp
},a\right\}  +\min\left\{  g^{\flat},b\right\}  \right)  $. By Lemma
\ref{lem2}, we have
\begin{align*}
\left(  \left(  f^{\sharp}\right)  _{a}^{\sharp\sharp}\right)  ^{\sharp}  &
=f^{\sharp\sharp}=f\\
\left(  \left(  g^{\flat}\right)  _{b}^{\flat\flat}\right)  ^{\flat}  &
=g^{\flat\flat}=g
\end{align*}
the last equalities occuring because $f$ is $u$-convex and $g$ is $v$-concave.
Taking conjugates in formula (\ref{qe}), we get $g\leq\bar{p}^{\flat}$ and
$f\geq\bar{p}^{\sharp}.$Substituting in the integral, we get:%
\[
\int_{X}\bar{p}^{\sharp}\left(  x\right)  d\mu-\int_{Y}\bar{p}^{\flat}\left(
y\right)  d\nu\leq\int_{X}f\left(  x\right)  d\mu-\int_{Y}g\left(  y\right)
d\nu
\]
and hence, by formula (\ref{oq}):%
\[
\int_{X}\bar{p}^{\sharp}\left(  x\right)  d\mu-\int_{Y}\bar{p}^{\flat}\left(
y\right)  d\nu\leq\inf_{a\leq p\leq b}\left[  \int_{X}p^{\sharp}\left(
x\right)  d\mu-\int_{Y}p^{\flat}\left(  y\right)  d\nu\right]
\]

Since $\bar{p}$ is admissible, $\bar{p}$ must be a minimizer, and the result follows.
\end{proof}

The proof indicates that uniqueness is not to be expected. The following
result is the Non-Uniqueness Theorem for prices:

\begin{proposition}
\label{p11}Let $p$ be a solution of problem (P). Then $p_{a}^{\sharp\sharp}$
and $p_{b}^{\flat\flat}$ are also solutions. More generally, if $q$ is an
admissible price schedule such that:%
\[
p_{a}^{\sharp\sharp}\left(  z\right)  \leq q\left(  z\right)  \leq
p_{b}^{\flat\flat}\left(  z\right)  \ \ \ \forall z\in Z_{1}%
\]
then $q$ is a solution of problem (P).
\end{proposition}

\begin{proof}
From $p_{a}^{\sharp\sharp}\leq q\leq p_{b}^{\flat\flat}$, we deduce that
$p^{\flat}=\left(  p_{b}^{\flat\flat}\right)  ^{\flat}\leq q^{\flat}$ and that
$q^{\sharp}\leq\left(  p_{a}^{\sharp\sharp}\right)  ^{\sharp}=p^{\sharp}$.
Substituting into the integral, we get:%
\[
\int_{X}q^{\sharp}\left(  x\right)  d\mu-\int_{Y}q^{\flat}\left(  y\right)
d\nu\leq\int_{X}p^{\sharp}\left(  x\right)  d\mu-\int_{Y}p^{\flat}\left(
y\right)  d\nu=\inf\left(  P\right)
\]
and since $q$ is admissible, it must be a minimizer.
\end{proof}

\begin{corollary}
Let $p$ be a solution of problem (P). Then $p^{\sharp}=\left(  p_{b}%
^{\flat\flat}\right)  ^{\sharp}$, $\mu$-almost everywhere, and $p^{\flat
}=\left(  p_{a}^{\sharp\sharp}\right)  ^{\flat}$, $\nu$-almost everywhere.
\end{corollary}

\begin{proof}
By the preceding Proposition, $p_{b}^{\flat\flat}$ is a solution of problem
(P), so that $I\left(  p_{b}^{\flat\flat}\right)  =I\left(  p\right)  .
$Substituting in the integrals, we get:
\[
\int_{X}\left(  p_{b}^{\flat\flat}\right)  ^{\sharp}d\mu-\int_{Y}\left(
p_{b}^{\flat\flat}\right)  ^{\flat}d\nu=\int_{X}p^{\sharp}\left(  x\right)
d\mu-\int_{Y}p^{\flat}\left(  y\right)  d\nu
\]
and since $\left(  p_{b}^{\flat\flat}\right)  ^{\flat}=p^{\flat}$, this
reduces to:%
\[
\int_{X}\left(  p_{b}^{\flat\flat}\right)  ^{\sharp}d\mu=\int_{X}p^{\sharp
}\left(  x\right)  d\mu
\]

Since $p_{b}^{\flat\flat}\leq p$, we have $\left(  p_{b}^{\flat\flat}\right)
^{\sharp}\geq p^{\sharp}$, and since the integrals are equal, it follows that
$p^{\sharp}=\left(  p_{b}^{\flat\flat}\right)  ^{\sharp}$, $\mu$-a.e. The same
argument shows that $p^{\flat}=\left(  p_{a}^{\sharp\sharp}\right)  ^{\flat}$,
$\nu$-a.e.
\end{proof}

\begin{corollary}
Let $p$ be a solution of problem (P). Then, for $\mu$-almost every $x$ in $X
$, there is a point $z\in D\left(  x\right)  $ such that $p\left(  z\right)
=p_{b}^{\flat\flat}\left(  z\right)  ,$ and for $\nu$-almost every $y$ in $Y
$, there is a point $z\in S\left(  y\right)  $ such that $p\left(  z\right)
=p_{a}^{\sharp\sharp}\left(  z\right)  $
\end{corollary}

\begin{proof}
Fix an $x$\ such that $p^{\sharp}\left(  x\right)  =\left(  p_{b}^{\flat\flat
}\right)  ^{\sharp}\left(  x\right)  $ and consider the functions $\varphi$
and $\psi$ defined by $\varphi\left(  z\right)  =u\left(  x,z\right)
-p\left(  z\right)  $ and $\psi\left(  z\right)  =u\left(  x,z\right)
-p_{b}^{\flat\flat}\left(  z\right)  $. We have $\varphi\geq\psi$, and
$\max\varphi=\max\psi$. So there must be a point $\bar{z}$ such that
$\max\varphi=\max\psi=\varphi\left(  \bar{z}\right)  =\psi\left(  \bar
{z}\right)  .$The result follows.
\end{proof}

Note that we already have $p\left(  z\right)  =p^{\sharp\sharp}\left(
z\right)  $ for every $z\in D\left(  x\right)  $, and $p\left(  z\right)
=p^{\flat\flat}\left(  z\right)  $ for every $z\in S\left(  y\right)  $

\subsection{The dual problem:\ optimality conditions}

Recall that we have defined a map $I:\mathcal{K}\left(  Z\right)  \rightarrow
R$ by:%
\[
I\left(  p\right)  =\int_{X}p^{\sharp}\left(  x\right)  d\mu-\int_{Y}p^{\flat
}\left(  y\right)  d\nu
\]

We have checked that the function $I$ is convex. It is easily seen to be
continuous: if $p_{n}\rightarrow p$ uniformly on $Z$, then $p_{n}^{\sharp
}\rightarrow p^{\sharp}$ uniformly on $X$ and $p_{n}^{\flat}\rightarrow
p^{\flat}$ uniformly on $Y$. On the other hand, the set$\ \mathcal{A}$ is
non-empty, convex and closed in $\mathcal{K}\left(  Z\right)  $. This means
that the constraint qualification conditions hold in problem (P): a necessary
and sufficient condition for $\bar{p}$ to be optimal is that:%
\begin{equation}
0\in\partial I\left(  \bar{p}\right)  +N_{\mathcal{A}}\left(  \bar{p}\right)
\label{oc}%
\end{equation}
where $\partial I\left(  \bar{p}\right)  $ is the subgradient of $I$ at
$\bar{p}$ in the sense of convex analysis, and $N_{\mathcal{A}}\left(  \bar
{p}\right)  $ is the normal cone to $\mathcal{A}$ at $\bar{p}$. All we have to
do now is to compute both of them.

\subsubsection{Computing $\partial I\left(  p\right)  $}

\begin{lemma}
\label{l71}Let $p\in\mathcal{K}\left(  Z\right)  $\ and $\varphi\in
\mathcal{K}\left(  Z\right)  $. Then, for every $x\in X$ and every $y\in Y$,
we have:%
\begin{align*}
\lim_{\substack{h\rightarrow0\\h>0}}\frac{1}{h}\left[  \left(  p+h\varphi
\right)  ^{\sharp}\left(  x\right)  -p^{\sharp}\left(  x\right)  \right]   &
=-\min\left\{  \varphi\left(  z\right)  \ |\ z\in D\left(  x\right)  \right\}
\\
\lim_{\substack{h\rightarrow0\\h>0}}\frac{1}{h}\left[  \left(  p+h\varphi
\right)  ^{\flat}\left(  y\right)  -p^{\flat}\left(  y\right)  \right]   &
=-\max\left\{  \varphi\left(  z\right)  \ |\ z\in S\left(  y\right)  \right\}
\end{align*}

\end{lemma}

\begin{proof}
Let us prove the second equality; the first one is derived in a similar way.
Take $z\in S_{p}\left(  y\right)  $ and $z_{h}\in S_{p+h\varphi}\left(
y\right)  $. From the definition of $S_{p}\left(  y\right)  $ and
$S_{p+h\varphi}\left(  y\right)  $, we have:%
\begin{align*}
v\left(  y,z_{h}\right)  -p\left(  z_{h}\right)   &  \geq p^{\flat}\left(
y\right)  =v\left(  y,z\right)  -p\left(  z\right) \\
v\left(  y,z\right)  -p\left(  z\right)  -h\varphi\left(  z\right)   &
\geq\left(  p+h\varphi\right)  ^{\flat}\left(  y\right)  =v\left(
y,z_{h}\right)  -p\left(  z_{h}\right)  -h\varphi\left(  z_{h}\right)
\end{align*}
Substracting, we find that:%
\begin{equation}
-h\varphi\left(  z\right)  \geq\left(  p+h\varphi\right)  ^{\flat}\left(
y\right)  -p^{\flat}\left(  y\right)  \geq-h\varphi\left(  z_{h}\right)
\label{ineq}%
\end{equation}
Since $z$ is an arbitrary point in $S_{p}\left(  y\right)  $, we can take it
to be the minimizer on the left-hand side, and this inequality becomes:%
\[
-h\max\left\{  \varphi\left(  z\right)  \ |\ z\in S_{p}\left(  y\right)
\right\}  \geq\left(  p+h\varphi\right)  ^{\flat}\left(  y\right)  -p^{\flat
}\left(  y\right)  \geq-h\varphi\left(  z_{h}\right)
\]

Now let $h\rightarrow0$. The family $z_{h}\in S_{p+h\varphi}\left(  y\right)
$ must have cluster points, because $Z$ is compact, and any cluster point
$\bar{z}$ must belong to $S_{p}\left(  y\right)  $. Taking limits in
inequality (\ref{ineq}), we find that, for some $\bar{z}\in S_{p}\left(
y\right)  $:%
\begin{equation}
-\max\left\{  \varphi\left(  z\right)  \ |\ z\in S_{p}\left(  y\right)
\right\}  \geq\lim_{\substack{h\rightarrow0\\h>0}}\frac{1}{h}\left[  \left(
p+h\varphi\right)  ^{\flat}\left(  y\right)  -p^{\flat}\left(  y\right)
\right]  \geq-\varphi\left(  \bar{z}\right)  \label{gh}%
\end{equation}
and the result follows.
\end{proof}

Because of inequality (\ref{gh}), we can apply the Lebesgue convergence
theorem, and we get:%
\begin{equation}
\lim_{\substack{h\rightarrow0\\h>0}}\frac{1}{h}\left[  I\left(  p+h\varphi
\right)  -I\left(  p\right)  \right]  =\int_{Y}\max\left\{  \varphi\left(
z\right)  \ |\ z\in S\left(  y\right)  \right\}  d\nu-\int_{X}\min\left\{
\varphi\left(  z\right)  \ |\ z\in D\left(  x\right)  \right\}  d\mu\label{k1}%
\end{equation}

We now work on the right-hand side of formula (\ref{k1}). Define $B\left(
X,D\right)  $ to be the set of all Borel maps $d:X\rightarrow Z$ such that
$d\left(  x\right)  \in D\left(  x\right)  $ for every $x$. Similarly,
$B\left(  Y,S\right)  $ is the set of all Borel maps $s:Y\rightarrow Z$ such
that $s\left(  y\right)  \in S\left(  y\right)  $ for every $y$.

\begin{lemma}
\label{l8} For every $\varphi\in\mathcal{C}\left(  Z\right)  $, we have:%
\begin{align}
\int_{X}\min\left\{  \varphi\left(  z\right)  \ |\ z\in D\left(  x\right)
\right\}  d\mu &  =\min\left\{  \int_{X}\varphi\left(  d\left(  x\right)
\right)  d\mu\ |\ d\in B\left(  X,D\right)  \right\} \label{ac}\\
\int_{Y}\max\left\{  \varphi\left(  z\right)  \ |\ z\in S\left(  y\right)
\right\}  d\nu &  =\max\left\{  \int_{Y}\varphi\left(  s\left(  y\right)
\right)  d\mu\ |\ s\in B\left(  Y,S\right)  \right\}  \label{ad}%
\end{align}

\end{lemma}

\begin{proof}
Given $\varphi\in\mathcal{C}\left(  Z\right)  $, the multivalued maps
$\Gamma_{1}$ and $\Gamma_{2}$ defined by:%
\begin{align*}
\Gamma_{1}\left(  x\right)   &  =\arg\min\left\{  \varphi\left(  z\right)
\ |\ z\in D\left(  x\right)  \right\} \\
\Gamma_{2}\left(  y\right)   &  =\arg\max\left\{  \varphi\left(  z\right)
\ |\ z\in S\left(  y\right)  \right\}
\end{align*}
have compact graph. Formulas (\ref{ac}) and (\ref{ad}) then follow from a
standard measurable selection theorem.
\end{proof}

Define $\mathcal{M}_{+}(X,D)$ to be the set of all demand distributions, that
is, the set of all positive measures $\alpha_{X\times Z}$ on $X\times Z$ which
are carried by the graph of $D$ and which have $\mu$ as marginal:%
\[
\alpha_{X}=\mu
\]

Recall that $\alpha_{Z}\in$ $\mathcal{M}_{+}(Z)$ denotes the second marginal
of $\alpha_{X\times Z}$.

\begin{lemma}
\label{l10}For every $\varphi\in\mathcal{K}\left(  Z\right)  $, we have:
\begin{equation}
\int_{X}\min\left\{  \varphi\left(  z\right)  \ |\ z\in D\left(  x\right)
\right\}  d\mu=\min\left\{  \int_{Z}\varphi d\alpha_{Z}\ |\ \alpha_{X\times
Z}\in\mathcal{M}_{+}(X,D)\right\}  \label{k3}%
\end{equation}

\end{lemma}

\begin{proof}
Let us investigate the right-hand side of formula (\ref{ac}). Let $f\in
B\left(  X,D\right)  $ be such that $\varphi\left(  f\left(  x\right)
\right)  =\min\left\{  \varphi\left(  z\right)  \ |\ z\in D\left(  x\right)
\right\}  $ for $\mu$-almost every $x$, and define $\gamma_{X\times Z}%
\in\mathcal{M}_{+}(X,D)$ by:
\[
\forall\psi\in\mathcal{K}\left(  X\times Z\right)  ,\ \ \ \int_{X\times Z}%
\psi\left(  x,z\right)  d\gamma_{X\times Z}=\int_{X}\psi\left(  x,f\left(
x\right)  \right)  d\mu
\]

Clearly:%
\begin{align*}
\min\left\{  \int_{X}\varphi\left(  d\left(  x\right)  \right)  d\mu\ |\ d\in
B\left(  X,D\right)  \right\}   &  =\int_{X}\varphi\left(  f\left(  x\right)
\right)  d\mu\ \\
&  =\int_{X\times Z}\varphi d\gamma_{X\times Z}\ =\int_{Z}\varphi d\gamma
_{Z}\\
&  \geq\min\left\{  \int_{Z}\varphi d\alpha_{Z}\ |\ \alpha_{X\times Z}%
\in\mathcal{M}_{+}(X,D)\right\}
\end{align*}

For the reverse inequality, we take any $\alpha_{X\times Z}\in\mathcal{M}%
_{+}(X,D)$. Taking conditional expectations, we have:%
\[
\ E_{x}^{\alpha}\left[  \varphi\right]  \geq\ \min\left\{  \varphi\left(
z\right)  \ |\ z\in D\left(  x\right)  \right\}
\]
and by integrating with respect to $\mu$, we get the desired result:%
\begin{align*}
\int_{Z}\varphi d\alpha_{Z}\  &  \geq\ \int_{X}\min\left\{  \varphi\left(
z\right)  \ |\ z\in D\left(  x\right)  \right\}  d\mu\\
&  =\min\left\{  \int_{X}\varphi\left(  z\right)  d\mu\ |\ z\in D\left(
x\right)  \right\} \\
&  =\min\left\{  \int_{X}\varphi\left(  d\left(  x\right)  \right)
d\mu\ |\ d\in B\left(  X,D\right)  \right\}
\end{align*}

\end{proof}

Considering the set $\mathcal{M}_{+}(Y,S)$ of supply distributions, we get
similar results:%
\begin{equation}
\int_{Y}\max\left\{  \varphi\left(  z\right)  \ |\ z\in S\left(  y\right)
\right\}  d\nu=\max\left\{  \int_{Z}\varphi d\beta_{Z}\ |\ \beta_{Y\times
Z}\in\mathcal{M}_{+}(Y,S)\right\}  \label{k4}%
\end{equation}

Writing formulas (\ref{k3}) and (\ref{k4}) in formula (\ref{k1}), we get:%

\begin{align*}
&  \lim_{\substack{h\rightarrow0\\h>0}}\frac{1}{h}\left[  I\left(
p+h\varphi\right)  -I\left(  p\right)  \right] \\
&  =\max\left\{  \int_{Z}\varphi d\beta_{Z}\ |\ \beta_{Y\times Z}%
\in\mathcal{M}_{+}(Y,S)\right\}  -\min\left\{  \int_{Z}\varphi d\alpha
_{Z}\ |\ \alpha_{X\times Z}\in\mathcal{M}_{+}(X,D)\right\} \\
&  =\max\left\{  \int_{Z}\varphi d\beta_{Z}-\int_{Z}\varphi d\alpha
_{Z}\ |\ \ \beta_{Y\times Z}\in\mathcal{M}_{+}(Y,S),\ \alpha_{X\times Z}%
\in\mathcal{M}_{+}(X,D)\right\}
\end{align*}

\begin{proposition}
\label{pk1} The subdifferential of $I$ at $p$ is given by:
\[
\partial I\left(  p\right)  =\left\{  \beta_{Z}-\alpha_{Z}\ |\ \beta_{Y\times
Z}\in\mathcal{M}_{+}(Y,S),\ \alpha_{X\times Z}\in\mathcal{M}_{+}(X,D)\right\}
\]

\end{proposition}

\begin{proof}
Take $\lambda\in\mathcal{M}\left(  Z\right)  =\mathcal{M}_{b}\left(  Z\right)
$. By definition of the subgradient, $\lambda\in\partial I\left(  p\right)  $
if and only if, for every $\varphi\in\mathcal{K}\left(  Z\right)  $ and $h>0,$
we have:
\[
I\left(  p+h\varphi\right)  \geq I\left(  p\right)  +h\int_{Z}\varphi d\lambda
\]

Since $I$ is convex, this is equivalent to:%
\[
\lim_{\substack{h\rightarrow0\\h>0}}\frac{1}{h}\left[  I\left(  p+h\varphi
\right)  -I\left(  p\right)  \right]  \geq\ \int_{Z}\varphi d\lambda
\]

Because of formula (\ref{k1}), this is equivalent to:%
\[
\max\left\{  \int_{Z}\varphi d\beta_{Z}-\int_{Z}\varphi d\alpha_{Z}%
\ |\ \beta_{Y\times Z}\in\mathcal{M}_{+}(Y,S),\ \alpha_{X\times Z}%
\in\mathcal{M}_{+}(X,D)\right\}  \ \geq\ \int_{Z}\varphi d\lambda
\]

This means that $\lambda$ belongs to the closed convex set:
\[
\left\{  \beta_{Z}-\alpha_{Z}\ |\ \beta_{Y\times Z}\in\mathcal{M}%
_{+}(Y,S),\ \alpha_{X\times Z}\in\mathcal{M}_{+}(X,D)\right\}
\]

\end{proof}

\subsubsection{Computing $N_{\mathcal{A}}\left(  p\right)  $}

Take $\lambda\in\mathcal{M}\left(  Z\right)  =\mathcal{M}_{b}\left(  Z\right)
$. By definition, $\lambda\in N_{\mathcal{A}}\left(  p\right)  $ if and only
if, for every $q\in\mathcal{A}$ $,$ we have:
\[
\int_{Z}\left(  q-p\right)  d\lambda\ \leq0
\]

Since $q\left(  \varnothing_{d}\right)  =p\left(  \varnothing_{d}\right)  =0$
and $q\left(  \varnothing_{s}\right)  =p\left(  \varnothing_{s}\right)  =0$
for every $q\in\mathcal{A}$, this condition is equivalent to:%
\begin{equation}
\int_{Z_{0}}\left(  q-p\right)  d\lambda\ \leq0 \label{g22}%
\end{equation}

To interpret this condition, we need some notation. Set:%
\begin{align*}
Z^{b}  &  =\left\{  z\ \in Z_{0}\ |\ a\left(  z\right)  <\ p\left(  z\right)
=b\left(  z\right)  \right\} \\
Z_{a}^{b}  &  =\left\{  z\ \in Z_{0}\ |\ a\left(  z\right)  <p\left(
z\right)  <b\left(  z\right)  \right\} \\
Z_{a}  &  =\left\{  z\ \in Z_{0}\ |\ a\left(  z\right)  =p\left(  z\right)
<b\left(  z\right)  \right\} \\
M  &  =\left\{  z\ \in Z_{0}\ |\ a\left(  z\right)  =p\left(  z\right)
=b\left(  z\right)  \right\} \\
N  &  =\left\{  z\ \in Z_{0}\ |\ a\left(  z\right)  >b\left(  z\right)
\right\}
\end{align*}
so that we have a partition of $Z_{0}$ into subsets $Z_{0}=Z_{a}\cup Z_{a}%
^{b}\cup Z^{b}\cup M\cup N$, where $Z_{a}\cup Z_{a}^{b}\cup Z^{b}\cup M=Z_{1}%
$, the set of marketable qualities.

Denote by $\lambda^{b},\ \lambda_{a}^{b},\ \lambda_{a},\ \lambda_{M}%
,\ \lambda_{N}$ the restrictions of $\lambda$ to $Z^{b},\ Z_{a}^{b}%
,\ Z_{a},\ Z_{M},\ Z_{N}~$respectively. Note that since $\lambda$ was a
bounded measure, so are $\lambda^{b},\ \lambda_{a}^{b},\ \lambda_{a}%
$\ $,\ \lambda_{M}$ and $\lambda_{N}$. Condition (\ref{g22}) is equivalent to
the following:%
\begin{equation}
\lambda^{b}\geq0,\ \lambda_{a}^{b}=0,~\lambda_{a}\leq0,\lambda_{N}=0
\label{g2}%
\end{equation}

\subsubsection{Concluding the proof.}

Let $\bar{p}$ be a solution of problem (P). By condition (\ref{oc}), we have
$0\in\partial I\left(  \bar{p}\right)  +N_{\mathcal{A}}\left(  \bar{p}\right)
$. By Proposition \ref{pk1} , this means that there exists $\beta_{Y\times
Z}\in\mathcal{M}_{+}(Y,S)$, $\alpha_{X\times Z}\in\mathcal{M}_{+}(X,D)$ and
$\lambda\in\mathcal{M}\left(  Z\right)  $ satisfying (\ref{g2}) such that
$\alpha_{Z}-\beta_{Z}=\lambda$.

In other words, the restriction of $\alpha_{Z}-\beta_{Z}$ to $Z^{b}%
,\ Z_{a}^{b},\ Z_{a}$ respectively are positive, zero and negative:%
\begin{align}
\alpha_{Z}  &  \geq\beta_{Z}\text{ \ on }Z^{b}\label{gh1}\\
\alpha_{Z}  &  =\beta_{Z}\text{ \ on }Z_{a}^{b}\label{gh2}\\
\alpha_{Z}  &  \leq\beta_{Z}\text{\ on }Z_{a}\label{gh3}\\
\alpha_{Z}  &  =\beta_{Z}\text{ \ on }N \label{gh4}%
\end{align}

There is no condition on the restriction of $\alpha_{Z}$ or $\beta_{Z}$ to
$\left\{  \varnothing_{d}\right\}  $,$\ \left\{  \varnothing_{s}\right\}  $ or
$M$. Since $P_{x}^{\alpha}$ is carried by $D\left(  x\right)  $, we must have
$P_{x}^{\alpha}\left(  z\right)  =0$ whenever $z\notin D\left(  x\right)  $,
which certainly is the case when $p\left(  z\right)  >b\left(  z\right)  $.
Similarly, $P_{y}^{\beta}\left(  z\right)  =0$ when $p\left(  z\right)
<a\left(  z\right)  $. If $z\in N$, either $p\left(  z\right)  >b\left(
z\right)  $ or $p\left(  z\right)  <a\left(  z\right)  $, so either
$P_{x}^{\alpha}\left(  z\right)  =0$ or $P_{y}^{\beta}\left(  z\right)  =0$.
The condition $\alpha_{Z}=\beta_{Z}$ on $N$ then implies that:%
\[
\alpha_{Z}=\beta_{Z}=0\text{ \ on }N
\]

We will now show that there exists $\alpha_{X\times Z}^{\prime}$
$\in\mathcal{M}_{+}(X,D)$ and $\beta_{Y\times Z}^{\prime}\in\mathcal{M}%
_{+}(Y,S)$ such that $\alpha_{Z_{0}}^{\prime}=\beta_{Z_{0}}^{\prime}$ . This
will be done by suitably modifying $\alpha_{X\times Z}$ and $\beta_{Y\times
Z}$ on the subsets $Z^{b}$ and $Z_{a}$ (note that they are both subsets of
$Z_{0}$). In the sequel, we will denote by $\alpha_{X\times A}$ (resp.
$\beta_{Y\times B}$) the restriction of $\alpha_{X\times Z}$ (resp.
$\beta_{Y\times Z}$) to $X\times A$ (resp. $Y\times B$), for $A\subset X$
(resp. $B\subset Y$), and by $\alpha_{A}$ (resp. $\beta_{B}$) the marginal on
$A$ (resp. $B$).

On $X\times Z^{b}$, we have, by:%
\[
\alpha_{X\times Z^{b}}=\int_{Z}P_{z}^{\alpha}d\alpha_{Z^{b}}\text{ \ and
\ }\beta_{X\times Z^{b}}=\int_{Z}P_{z}^{\beta}d\beta_{Z^{b}}%
\]
with $\alpha_{Z^{b}}\geq\beta_{Z^{b}}$ by (\ref{gh1}). Define $\alpha_{X\times
Z}^{\prime}$ by:
\begin{align*}
\alpha_{X\times Z^{b}}^{\prime}  &  =\int_{Z}P_{z}^{\alpha}d\beta_{Z^{b}}\\
\alpha^{\prime}\left(  X\times\left\{  \varnothing_{d}\right\}  \right)   &
=\alpha\left(  X\times\left\{  \varnothing_{d}\right\}  \right)
+\alpha\left(  Z^{b}\right)  -\beta\left(  Z^{b}\right) \\
\alpha_{X\times\left(  Z-Z^{b}\cup\left\{  \varnothing_{d}\right\}  \right)
}^{\prime}  &  =\alpha_{X\times\left(  Z-Z^{b}\cup\left\{  \varnothing
_{d}\right\}  \right)  }%
\end{align*}

Clearly $\alpha_{X\times Z}^{\prime}$ is a positive measure. It follows from
the first equation that $\alpha_{Z^{b}}^{\prime}=\beta_{Z^{b}}$, and from the
second that $\alpha_{x}^{\prime}=\alpha_{X}=\mu$. It remains to check that
$\alpha_{X\times Z}^{\prime}$ $\in\mathcal{M}_{+}(X,D)$. We already know that
$\alpha_{X\times Z}$ $\in\mathcal{M}_{+}(X,D)$, meaning that for
$P_{x}^{\alpha}\left[  D\left(  x\right)  \right]  =1$ for $\mu$-a.e. $x$, and
it differs from $\alpha_{X\times Z}^{\prime}$ only in the region where $z\in
Z^{b}$ or $z=\varnothing_{d}$. If $D\left(  x\right)  \cap Z^{b}=\varnothing$
then $P_{x}^{\alpha}\left[  D\left(  x\right)  \right]  =1$ as well. If
$D\left(  x\right)  $ intersects $Z^{b}$, so that $z\in Z^{b}\cap D\left(
x\right)  $, then consumer $x$ is paying the highest bid price for $z$, and so
he must be indifferent between $z$ and $\varnothing_{d}$; this shows that
$\varnothing_{d}$ also belongs to $D\left(  x\right)  $. In the new allocation
$\alpha_{X\times Z}^{\prime}$, some of the demand may be transfered from
$Z^{b}\cap D\left(  x\right)  $ to $\varnothing_{d}$ with positive
probability, but this redistribution occurs within $D\left(  x\right)  $ and
does not affect the total probability, so that $P_{x}^{\alpha^{\prime}}\left[
D\left(  x\right)  \right]  =1.$

In words, for every quality $z$ where the highest bid price is paid, we clear
the market by letting some of the demand go unsatisfied: all producers $y$
have sold, but there is total quantity $\alpha\left(  Z^{b}\right)
-\beta\left(  Z^{b}\right)  $ of potential buyers which are thrown out of the
market. However, they don't care, because the price asked is the highest bid
price, and they are indifferent between buying or nor.

We then shift some of the supply to $\varnothing_{s}$, as we did for the
demand. We end up with $\alpha_{X\times Z}^{\prime}\in\mathcal{M}_{+}(X,D)$
and $\beta_{Y\times Z}^{\prime}\in\mathcal{M}_{+}(Y,S)$ which satisfy the
conclusions of the Existence Theorem.

\section{Remaining proofs\label{sec21}}

\subsection{Pareto optimality of equilibrium allocations}

With every pair of demand and supply distributions, $\alpha_{X\times
Z}^{\prime}\in\mathcal{M}_{+}(X,D)$ and $\beta_{Y\times Z}^{\prime}%
\in\mathcal{M}_{+}(Y,S)$, we associate the number:%

\begin{align*}
J\left(  \alpha_{X\times Z}^{\prime},\beta_{Y\times Z}^{\prime}\right)   &
=\int_{X\times Z}u\left(  x,z\right)  d\alpha_{X\times Z}^{\prime}%
-\int_{Y\times Z}v\left(  y,z\right)  d\beta_{Y\times Z}^{\prime}\\
&  =\int_{X}E_{x}^{\alpha^{\prime}}\left[  u\left(  x,z\right)  \right]
d\mu\left(  x\right)  -\int_{Y}E_{y}^{\beta^{\prime}}\left[  v\left(
y,z\right)  \right]  d\nu\left(  y\right)
\end{align*}

Assume that $\alpha_{Z_{0}}^{\prime}=\beta_{Z_{0}}^{\prime}$.We claim that:%
\begin{equation}
\int_{X}E_{x}^{\alpha^{\prime}}\left[  p\left(  z\right)  \right]  d\mu\left(
x\right)  -\int_{Y}E_{y}^{\beta^{\prime}}\left[  p\left(  z\right)  \right]
d\nu\left(  y\right)  =0 \label{cm1}%
\end{equation}

Indeed, the left-hand side can be written as:%
\[
\left(  \int_{Z_{0}}p\left(  z\right)  d\alpha_{Z}^{\prime}-\int_{Z_{0}%
}p\left(  z\right)  d\beta_{Z}^{\prime}\right)  +p\left(  \varnothing
_{d}\right)  \left(  \alpha_{Z}^{\prime}\left[  \varnothing_{d}\right]
-\beta_{Z}^{\prime}\left[  \varnothing_{d}\right]  \right)  +p\left(
\varnothing_{s}\right)  \left(  \alpha_{Z}^{\prime}\left[  \varnothing
_{s}\right]  -\beta_{Z}^{\prime}\left[  \varnothing_{s}\right]  \right)
\]

The first term vanishes because $\alpha_{Z_{0}}^{\prime}=\beta_{Z_{0}}%
^{\prime}$, and the two next terms vanish because $p\left(  \varnothing
_{d}\right)  =p\left(  \varnothing_{s}\right)  =0$.

Substracting (\ref{cm1}) from $J$, we get:%
\begin{equation}
J\left(  \alpha_{X\times Z}^{\prime},\beta_{Y\times Z}^{\prime}\right)
=\int_{X}E_{x}^{\alpha^{\prime}}\left[  u\left(  x,z\right)  -p\left(
z\right)  \right]  d\mu\left(  x\right)  -\int_{Y}E_{y}^{\beta^{\prime}%
}\left[  v\left(  y,z\right)  -p\left(  z\right)  \right]  d\nu\left(
y\right)  \label{cm3}%
\end{equation}

By Fenchel's inequality, $\left(  u\left(  x,z\right)  -p\left(  z\right)
\right)  \leq p^{\sharp}\left(  x\right)  $ for all $z\in Z$. Taking
expectations with respect to the probability $P_{x}^{\alpha^{\prime}}$, we
get:%
\begin{equation}
E_{x}^{\alpha^{\prime}}\left[  u\left(  x,z\right)  -p\left(  z\right)
\right]  \leq p^{\sharp}\left(  x\right)  \label{cm5}%
\end{equation}
with equality if and only if $u\left(  x,z\right)  -p\left(  z\right)
=p^{\sharp}\left(  x\right)  $ (in other words, $z\in D\left(  x\right)  $)
for $P_{x}^{\alpha^{\prime}}$-almost every $z\in Z$. Similarly, we have:%
\begin{equation}
E_{y}^{\beta^{\prime}}\left[  v\left(  y,z\right)  -p\left(  z\right)
\right]  \geq p^{\flat}\left(  y\right)  \label{cm6}%
\end{equation}
with equality if and only if $v\left(  y,z\right)  -p\left(  z\right)
=p^{\flat}\left(  y\right)  $ (in other words, $z\in S\left(  y\right)  $) for
$P_{y}^{\beta^{\prime}}$-almost every $z\in Z$. Writing this in (\ref{cm3}),
and treating the second term in the same way, we get:%
\begin{equation}
J\left(  \alpha_{X\times Z}^{\prime},\beta_{Y\times Z}^{\prime}\right)
\leq\int_{X}p^{\sharp}\left(  x\right)  d\mu-\int_{Y}p^{\flat}\left(
y\right)  d\nu\label{cm4}%
\end{equation}

The right-hand side is equal to $J\left(  \alpha_{X\times Z},\beta_{Y\times
Z}\right)  $, for any equilibrium allocation $\left(  \alpha,\beta\right)  $.
This proves that equilibrium allocations solve the planner's problem, and as
such they are Pareto optimal.

\subsection{Uniqueness of equilibrium allocations}

Observe that equality holds in (\ref{cm4}) if and only if equality holds in
(\ref{cm5}) for $\mu$-almost every $x$, and equality holds in (\ref{cm6}) for
$\nu$-almost every $y$. This means that $P_{x}^{\alpha^{\prime}}\left[
D\left(  x\right)  \right]  =1$ for $\mu$-almost every $x$ and $P_{y}%
^{\beta^{\prime}}\left[  S\left(  y\right)  \right]  =1$ for $\nu$-almost
every $y$.

\subsection{Proof of Theorem \ref{thm7}}

Let $\left(  p,\alpha_{X\times Z},\beta_{Y\times Z}\right)  $ be an
equilibrium. By Rademacher's theorem, since $p^{\sharp}:X\rightarrow R$ is
Lipschitz, and $\mu$ is absolutely continuous with respect to the Lebesgue
measure, $p^{\sharp}$ is differentiable $\mu$-almost everywhere.

Consider the set $A=\left\{  x\ |\ p^{\sharp}\left(  x\right)  \geq0\right\}
$. Let $x\in A$ be a point where $p^{\sharp}$ is differentiable, with
derivative $D_{x}p^{\sharp}\left(  x\right)  $. Since $x$ is active or
indifferent, the set $D\left(  x\right)  \cap Z_{0}$ is non-empty, and we may
take some $z\in D\left(  x\right)  \cap Z_{0}$. Consider the function
$\varphi\left(  x^{\prime}\right)  =u\left(  x^{\prime},z\right)  -p\left(
z\right)  $. By proposition \ref{prop1}, since $D\left(  x\right)
\subset\partial p^{\sharp}\left(  x\right)  $, we have $\varphi\leq f$ and
$\varphi\left(  x\right)  =f\left(  x\right)  $, so that $\varphi$ and $f$
must have the same derivative at $x:$
\begin{equation}
D_{x}f\left(  x\right)  =D_{x}u\left(  x,z\right)  \label{eq10}%
\end{equation}

By condition (\ref{eq11}), this equation defines $z$ uniquely. In other words,
for $\mu$-almost every point $x\in A$, the set $D\left(  x\right)  \cap Z_{0}$
consists of one point only. Similarly, for $\nu$-almost every point $y\in
B=\left\{  y\ |\ p^{\flat}\left(  y\right)  \leq0\right\}  $, the set
$S\left(  y\right)  \cap Z_{0}$ consists of one point only. This is the
desired result.

\end{document}